\documentclass[a4paper,fleqn,usenatbib]{mnras}

\usepackage{newtxtext,newtxmath}

\usepackage[T1]{fontenc}
\usepackage{ae,aecompl}

\usepackage{graphicx}	
\usepackage{amsmath}	
\usepackage{amssymb}	

\newcommand{\msun}{M$_{\odot}$}

\newcommand{\kms}{km~s$^{-1}$}

\newcommand{\Ha}{H$\alpha$}
\newcommand{\Hb}{H$\beta$}

\newcommand{\HeI}{He\hspace{0.25em}{\sc i}}

\newcommand{\OI}{O\hspace{0.25em}{\sc i}}

\newcommand{\NaI}{Na\hspace{0.25em}{\sc i}}
\newcommand{\MgII}{Mg\hspace{0.25em}{\sc ii}}

\newcommand{\SiII}{Si\hspace{0.25em}{\sc ii}}

\newcommand{\CaII}{Ca\hspace{0.25em}{\sc ii}}

\newcommand{\FeII}{Fe\hspace{0.25em}{\sc ii}}

\newcommand{\Fefs}{$^{56}$Fe}
\newcommand{\Cofs}{$^{56}$Co}
\newcommand{\Nifs}{$^{56}$Ni}
\newcommand{\mej}{$M_\mathrm{ej}$}
\newcommand{\ek}{$E_\mathrm{k}$}

\newcommand{\lp}{$L_\mathrm{p}$}

\newcommand{\eom}{$E_\mathrm{k}/M_\mathrm{ej}$}

\newcommand{\mni}{$M_\mathrm{Ni}$}

\newcommand{\tp}{$t_\mathrm{p}$}

\newcommand{\mN}{$\left<N\right>$}
\newcommand{\R}{$f_\mathrm{em}/f_\mathrm{abs}$}

\newcommand{\mhe}{$M_\mathrm{He}$}
\newcommand{\mh}{$M_\mathrm{H}$}
\newcommand{\angstrom}{$\mbox{\normalfont\AA}$}

\title[A 22 \msun\ Case Study]{How much H and He is ``hidden" in SNe Ib/c? - II. Intermediate-mass objects: a 22\,\msun\ progenitor case study}

\author[Teffs et al.]{
Jacob Teffs$^{1}$,\thanks{j.j.teffs@ljmu.ac.uk}
Thomas Ertl$^{2}$,
Paolo Mazzali$^{1,2}$,
Stephan Hachinger$^{3}$,
\newauthor
H.-Thomas Janka$^{2}$
\\
$^{1}$Astrophysics Research Institute, Liverpool John Moores University, ic2, 146 Brownlow Hill, Liverpool L3 5RF, UK\\
$^{2}$Max-Planck Institut f\"ur Astrophysik, Karl-Schwarzschild-Str. 1, D-85741 Garching, Germany\\
$^{3}$Leibniz Supercomputing Centre (LRZ), Boltzmannstr. 1, D-85748 Garching, Germany
}

\date{Accepted XXX. Received YYY; in original form ZZZ}

\pubyear{2020}

\begin{document}
\label{firstpage}
\pagerange{\pageref{firstpage}--\pageref{lastpage}}
\maketitle

\begin{abstract}
Stripped envelope supernovae are a sub-class of core collapse supernovae showing several stages of H/He shell stripping that determines the class: H-free/He-poor SNe are classified as Type Ic, H-poor/He-rich are Type Ib, and H/He-rich are Type IIb. Stripping H/He with only stellar wind requires significantly higher mass loss rates than observed while binary-involved mass transfer may usually not strip enough to produce H/He free SNe. Type Ib/c SNe are sometimes found to include weak H/He transient lines as a product of a trace amount of H/He left over from stripping processes. The extent and mass of the H/He required to produce these lines is not well known. In this work, a 22 \msun\ progenitor model is stripped of the H/He shells in five steps prior to collapse and then exploded at four explosion energies. Requiring both optical and NIR \HeI\ lines for helium identification does not allow much He mass to be hidden in SE--SNE. Increasing the mass of He above the CO core delays the visibility of \OI\,7774 in early spectra. Our SN\,Ib-like models are capable of reproducing the spectral evolution of a set of observed SNe with reasonable estimated \ek\ accuracy. Our SN\,IIb-like models can partially reproduce low energy observed SN\,IIb, but we find no observed comparison for the SN\,IIb-like models with high \ek. 

\end{abstract}



\section{Introduction} \label{sec:intro} 
Stripped envelope supernovae (SE--SNe) are a sub-class of core collapse supernovae in which prior to core collapse, the star experiences significant mass loss during their evolution \citep{1997ARA&A..35..309F}.  This may occur through binary interaction or through periods of high mass loss, with rates of 10$^{-4}$ to 10$^{-5}$ \msun\ yr$^{-1}$ \citep{NOMOTO1995173}. If the hydrogen envelope is mostly stripped away the SN, is of Type IIb; if the entire hydrogen envelope and parts of the helium are stripped, the SN is of Type Ib; and if all of the hydrogen and most of the helium are stripped, a Type Ic results \citep{1995ApJ...450L..11F}. The sub-types of SE--SNe can be difficult to separate, with some SNe being classified early as a He-free Type Ic before transitioning into a Type Ib with weak He features. With no hydrogen envelope recombining to power a plateau-like phase typical of Type IIP \citep{1994ApJ...430..300E}, the primary source of luminosity in these SE--SNe is the radioactive decay of \Nifs\ and eventually \Cofs.

The extent and mechanism of the stripping of the H/He shells that result in observed SE--SNe is likely due to three primary mechanisms. For single star evolution, two mechanisms thought to strip H and He prior to collapse are high mass loss rates from wind (e.g. \citet{NOMOTO1995173}) or episodic mass loss during a luminous blue variable stage \citep{Smith2003}. However, stellar evolution models for single stars using these mechanisms are often not able to remove enough H/He to prevent the formation of \HeI\ or \Ha\ lines in the spectra to explain the diversity of the SE--SNe \citep{2012MNRAS.422...70H}. The luminous blue variable stage occurs in high mass progenitors but may not be able to explain low mass events like SN1994I \citep{Nomoto1994,2006MNRAS.369.1939S} and still may not strip enough He to be lower than suggested observed limits (e.g. \citet{2012MNRAS.422...70H}). For stars evolving in a binary, mass loss during a common envelope or Roche lobe phase can explain some SE--SNe \citep{NOMOTO1995173,Yoon_2010}, but can leave thin shells of H/He of low mass above the carbon and oxygen rich (CO) core. This can explain the sometimes observed weak He lines in Type Ic and weak H lines in Type Ib \citep{Branch_2002,Elmhamdi2006}. However, not all Type Ic SNe show weak or transient He lines and not all Type Ib SNe show weak or transient H lines \citep{Drout2016,10.1093/mnras/stx980}. 

The formation of He lines during a supernova requires significant deviations from a local thermodynamic equilibrium (LTE) state, with previous spectral simulations under LTE conditions artificially boosting the departure coefficients of helium by $\sim$10$^4$--10$^{10}$ in order to produce the observed \HeI\ lines in Type Ib SNe \citep{1987ApJ...317..355H,1991ApJ...383..308L}. The significant deviation from LTE conditions required for \HeI\ line formation can be achieved by the collisions of fast electrons resulting from Compton processes from $\gamma$-rays generated by the decay of \Nifs\ and \Cofs\ \citep{1988ApJ...335L..53G}. Nucleosynthesis during the core collapse process produces \Nifs\ in the Fe/Si rich inner core \citep{1995ApJS..101..181W}. In order for the \Nifs\ decay products to excite the He, either the \Nifs\ is mixed outward into the ejecta, the He is mixed inward, the expanding ejecta thins which reduces the $\gamma$-ray blocking, or a combination of the all three factors \citep{1990ApJ...361L..23S,1991ApJ...383..308L}. 

The mixing of \Nifs\ and other elements in the ejecta is often required to successfully model SNe but the mechanism and extent can be varied across different supernovae types as well as within a specific sub-class. Mixing this material often requires multi-dimensional hydrodynamics simulations and is likely a key factor in a successful core collapse explosion \citep{1995ApJ...450..830B,1996A&A...306..167J}. Observations of photospheric and nebular spectra suggest that the explosion mechanism of SE--SNe is likely asymmetric in nature \citep{Mazzali1284,2008ARA&A..46..433W,Tanaka_2009,2017MNRAS.469.1897S}. Using abundance tomography \citep{2005MNRAS.360.1231S} to model these events often requires some or significant amounts of mixing for a wide range of \ek\ and \mej\, \citep{1998Natur.395..672I,2004ApJ...614..858M,10.1093/mnras/stz1588}.

This work is a partial continuation of two previous works. First, \citet{2012MNRAS.422...70H} focused on a pair of models fitted to two lower mass SE--SNe events; the narrow lined SN\,Ic 1994I (e.g.\ \citet{1995ApJ...450L..11F}) and the SN\,IIb 2008ax (e.g.\ \citet{2011MNRAS.413.2140T}). These two SNe had \ek\ $\sim$1 foe with \mej\  $\sim$1 \msun\, and 2--5 \msun\, respectively, where 1 foe is 1$\times 10^{51}$ ergs. Using a series of H/He shells between these two models, an estimated mass limit in the line forming region required to produce a set of identifiable \HeI\ lines or \Ha\, in the spectra was found. 

Then, \citet{10.1093/mnras/staa123} narrowed the focus to a single 22\,\msun\ progenitor stripped of all H/He to explore explosion mixing and energy and the resulting Type Ic-like spectra. Using this model, we found that a combination of abundance mixing and \ek\ for this progenitor mass can reproduce the bulk spectral features of a wide range of SNe\,Ic, from Ic-BL such as 2002ap \citep{2002ApJ...572L..61M} to narrow lined SNe like 1994I. As \mej\ and \mni\ were kept fixed in that work, the resulting peak luminosity (\lp\,) from the synthetic light curve normally did not match despite similar light curve widths. However, we found that by comparing the spectral shape and colour, the approximate de--reddened line velocities, and the extent of line blending of peak or near-peak observed spectra to the synthetic spectra, we were able to estimate the \ek\ of several SNe\,Ic. We primarily focused on the near peak epochs, but line blending and features can change as a function of time. For a smaller set of observed SNe, we compared the spectral evolution to the spectra SNe of similar \ek\ and found good comparisons.  

Following the ground work set in the previous two papers, we consider an evolved stellar model with various stages of pre--core collapse stripping. We explore the effects of stripping on the resulting bolometric light curves and spectral evolution. The stellar model is stripped of some fraction of its hydrogen and helium outer shells in 5 stages until a bare CO core is left, with only trace amounts of atmospheric helium remaining. Each stripped model is exploded at four final explosion energies and three mixing approximations are used to expand the parameter space for comparisons to observed SE--SNe. We discuss the stripped models, the spectral synthesis, and the light curve calculation in Section \ref{sec:meth}. The spectral models for a fully stripped CO core with trace amounts of helium are discussed in Section \ref{sec:co_core}. The He-poor/rich models with no hydrogen are discussed in Section \ref{sec:he_mods} and the He--rich/H--poor models are discussed in Section \ref{sec:h_mods}. Some observed SNe are used as comparisons to the synthetic spectra in Sections \ref{sec:obs_Ib_sne} and \ref{sec:obs_IIb_sne}. In Section \ref{sec:disc} we discuss the results and in Section \ref{sec:conc}, we summarize our results and consider future extensions.  

\section{Methods} \label{sec:meth}
In the following sections we describe the methods used to generate the bolometric light curves and synthetic spectra. 

\subsection{Progenitor Models} \label{subsec:models}
The 22\,\msun\, progenitor model used here and in \citet{10.1093/mnras/staa123} is a modified version of the 22\,\msun\, non-rotating, solar metallicity stellar model generated by \citet{RevModPhys.74.1015}. The star is artificially stripped of H/He--rich material above the CO core in stages to produce possible SN IIb/Ib/Ic progenitors. This results in a bare CO core of 4--4.5 \msun\ with an increasing \mhe\ and \mh\ above it as less stripping is applied. Table \ref{tab:model_params} shows the resulting model parameters. 

\begin{table*} 
\caption{ The ejecta, nickel, helium, and hydrogen masses, the explosion energy, and the ratios \eom\ of all the models used in this work.}
\begin{tabular}{ccccccc}
    \hline
Model & \ek\ (foe)	&	\mej\,/\msun\, & \mni\,/\msun\,	& M$_{\text{H}}$/\msun\, & \mhe\,/\msun\, &	 \eom\, (foe/\msun\,)	\\
\hline
\hline
CO Core       & 1	& 4.04  & 0.096 & 0.0  & 0.018  & 0.248  \\
              & 3	& 4.24  & 0.147 & 0.0  & 0.018  & 0.708  \\
              & 5	& 4.39  & 0.186 & 0.0  & 0.018  & 1.139  \\
              & 8   & 4.54  & 0.225 & 0.0  & 0.020  & 1.762  \\
0.2 He        & 1	& 4.25  & 0.096 & 0.0  & 0.19    & 0.235  \\
              & 3	& 4.47  & 0.146 & 0.0  & 0.21    & 0.671  \\
              & 5	& 4.63  & 0.186 & 0.0  & 0.21    & 1.080  \\
              & 8   & 4.78  & 0.224 & 0.0  & 0.22    & 1.674  \\
0.5 He        & 1	& 4.57  & 0.096 & 0.0  & 0.51    & 0.219  \\
              & 3	& 4.78  & 0.147 & 0.0  & 0.51    & 0.628  \\
              & 5	& 4.94  & 0.186 & 0.0  & 0.51    & 1.012  \\
              & 8   & 5.08  & 0.225 & 0.0  & 0.51    & 1.575  \\
1.0 He        & 1	& 5.11  & 0.096 & 0.0  & 1.02    & 0.196  \\
              & 3	& 5.31  & 0.146 & 0.0  & 1.02    & 0.565  \\
              & 5	& 5.45  & 0.186 & 0.0  & 1.00    & 0.917  \\
              & 8   & 5.60  & 0.224 & 0.0  & 1.05    & 1.429  \\
1.3 He + 0.1 H & 1   & 5.55  & 0.097 & 0.11 & 1.36    & 0.180  \\
              & 3	& 5.76  & 0.146 & 0.11 & 1.36    & 0.521  \\
              & 5	& 5.93  & 0.186 & 0.11 & 1.36    & 0.843  \\
              & 8   & 6.06  & 0.224 & 0.11 & 1.36    & 1.320  \\
    \hline
\end{tabular}
\label{tab:model_params}
\end{table*}

The supernova explosion is simulated in a 1-D hydrodynamics code with neutrino transport called \textsc{Prometheus-HOTB} that is discussed in more detail by \citet{1996A&A...306..167J,2012ApJ...757...69U,ertl2016,ertl2020}. The neutrino heating used to generate the explosion is artificially enhanced in order to achieve a final \ek\ of 1, 3, 5, and 8 foe. This energy is deposited after core collapse and bounce in the surroundings of the newly formed neutron star over a timescale of several seconds. This is supposed to mimic, in a parametric way, the dynamical consequences of an engine that generates thermal energy by neutrinos or magnetic field effects, as required to explode the 22\,\msun\ star at the chosen \ek. In order to artificially generate this \ek, a larger mass of neutrino-heated ejecta is required. This results in the higher masses for the higher \ek\ models using the same stripping as seen in Table \ref{tab:model_params}. Each model, after energy injection, is evolved until the model has reached homologous expansion, such that $r \propto v t$, which occurs a few days after explosion. 

The explosive nucleosynthesis is followed using a 14-isotope nuclear network that tracks the bulk production of $\alpha$-elements up to \Nifs. This network includes a ``tracer" that accounts for iron-group nuclei in the slightly neutron-rich neutrino-heated ejecta, whose exact composition depends on the details of the neutrino transport \citep{ertl2016,2016ApJ...821...38S}. The \mni\ given in Table \ref{tab:model_params} is comprised of the \Nifs\ that is produced explosively and 50\%\ of the ``tracer" material. As the \ek\ increases, the ratio of \mni\ to  ``tracer" mass becomes smaller, so at higher energies, the \mni\ in Table \ref{tab:model_params} is dominated by the ``tracer" material. As such, the \mni\ used in this work should be treated as an approximate \mni\ value, and can vary by a non-negligible amount depending on the fraction of ``tracer" material considered to be \Nifs. Another side effect is that the innermost region is rich in He (approximately 0.1 - 0.4 \msun\, left over from $\alpha$-rich freezeout) and a mix of ``tracer" material and \Nifs. As this material's composition is uncertain, we also choose to replace the innermost \mhe\ with O as it is the most abundant element in CO cores and does not introduce new or overly strong lines, see \citet{10.1093/mnras/staa123} for a more complete discussion..

\subsection{Abundance Profiles} \label{subsec:abu_prof}
\begin{figure*}
        %
        \includegraphics[scale=.5,angle=0]{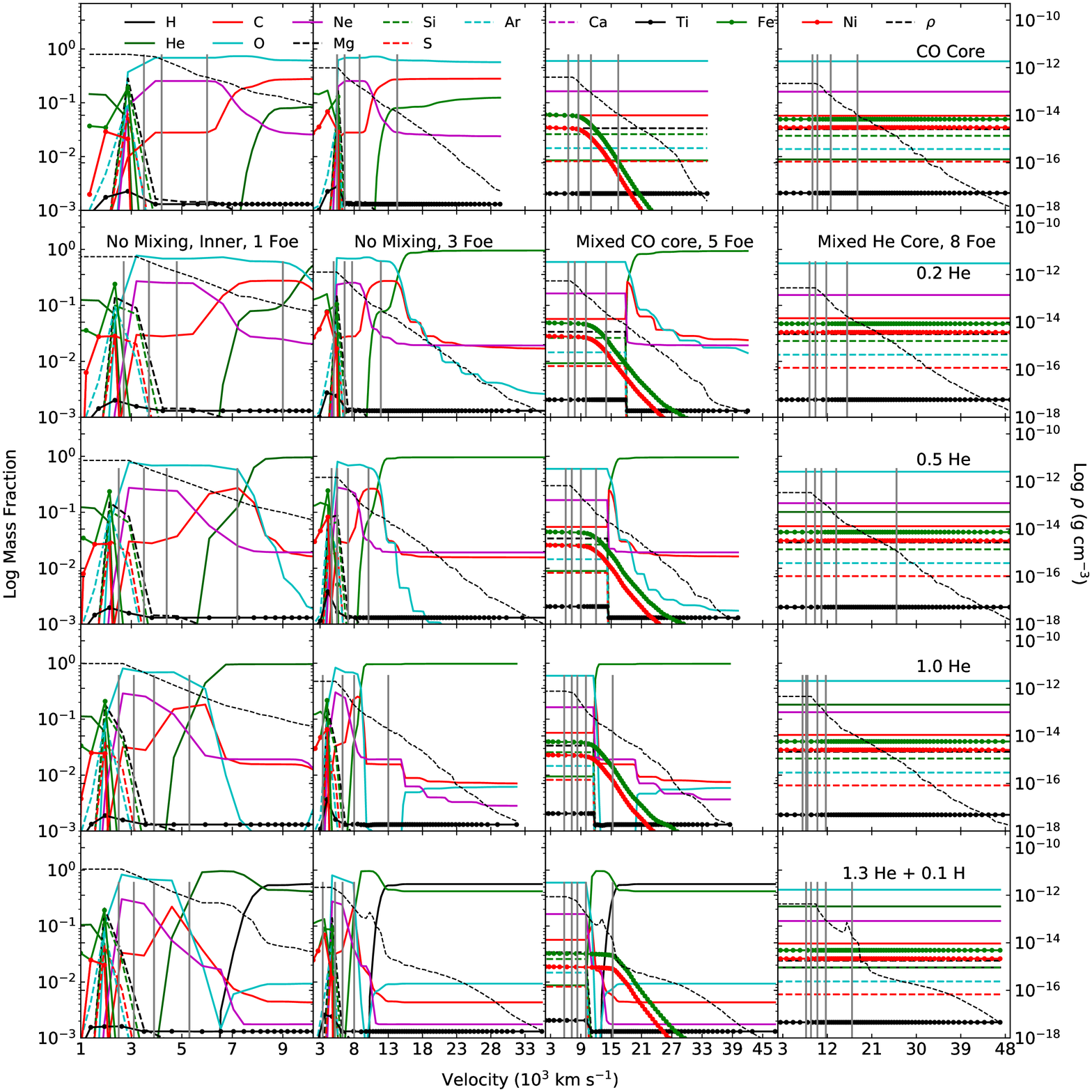}
        \caption{The abundance profiles of all 5 stripped models at 4 explosion energies and 3 mixing approximations. The first column is the innermost 10000 \kms\ of the unmixed models at 1 foe, the second is the unmixed models at 3 foe, the third is the mixed CO cores at 5 foe, and the fourth column is the mixed He cores at 8 foe.
        \label{fig:all_abund}}
\end{figure*}

Similar to \citet{10.1093/mnras/staa123}, we choose three mixing approximations to explore the effect of mixing on both the light curve and the synthetic spectral evolution. Our first mixing approximation, is the original, or untouched, model. For this mixing approximation, the abundance profiles are left completely untouched, but the innermost density is averaged, conserving total mass, following the methods described in \citet{10.1093/mnras/staa123}. This approximation is shown in the first two columns of Figures \ref{fig:all_abund} for  only the 1 and 3 foe cases. As the total combination of mixing, stripping, and energy results in 60 abundance profiles, we only show several select sets. The second mixing approximation is to mix only the CO core. This will be typically be called ``mixed CO core models" or models with a mixed CO core depending on the context. The gradual He shell stripping would then result in a series of lower and lower mass He shells on top of a fully mixed CO core. A running boxcar is used to mix out the \Nifs\ and Fe, resulting in a higher abundances in the core and lower in the He--rich material. This approximation is shown in the third column of Figure \ref{fig:all_abund} for the 5 foe case. The final mixing approximation is to completely mix the entire ejecta resulting in a uniform composition. This approximation will be referred to as ``mixed He core'' or models with fully mixed ejecta depending on the context. This approximation is shown in the fourth column of Figure \ref{fig:all_abund} for the 8 foe case.

\subsection{Light Curve and Spectral Synthesis Codes} \label{subsec:lc_specsyn}

After each explosion has reach homologous expansion, at $\sim 5$ days, we apply the mixing modifications discussed previously. The radioactive decay powered phase is calculated using our light curve code described in detail in \citet{1997A&A...328..203C}. The code calculates the emission and propagation of $\gamma$-rays and positrons produced by the decay of \Nifs, and subsequently \Cofs, into the homologously expanding ejecta using a Monte Carlo method. This code only treats the radioactively powered phase and does not treat shock breakout emission or other energy sources. We assume constant values for the positron and $\gamma$-ray opacities with values of 7\,cm$^2$g$^{-1}$ and 0.027\,cm$^2$g$^{-1}$, respectively \citep{1980PhDT.........1A}. The deposited energy is recycled into optical photons, which are also followed using a Monte Carlo scheme. The optical opacity is dependent on time and metallicity using a grey-atmosphere approximation that is designed to reproduce the dominance of line opacity in the ejecta \citep{2001ApJ...547..988M}. The ejecta are evolved as a function of time. The position of the photosphere is approximately determined by integrating inwards until the radius where an optical depth of $\tau$ $\ge$ 1 is found. The resulting photospheric velocity, combined with the luminosity, abundances, and epoch, are used as inputs in the spectral synthesis code.

The synthetic spectra are calculated using a Monte Carlo code discussed in detail in \citet{1993A&A...279..447M, 1999A&A...345..211L, 2000A&A...363..705M}. The code reads in the density and composition structure, the luminosity, and the velocity of the photosphere at a given epoch. As the photosphere recedes into the expanding ejecta, the near-photospheric region, where line formation is most likely to occur, can be characterized by a varying composition.  For the models in this work, the non-thermal contributions of H and/or He are explicitly taken into account as \mhe\ is large enough to reasonably expect the formation of \HeI\ lines in the spectra.

For all models, \tp\ is defined as the time when the bolometric light curve reaches maximum luminosity, or \lp. For simpler comparison among the models, the synthetic spectra for all energies and mixing approximations are generated at the same 7 epochs relative to \tp, rather than at epochs relative to the time of explosion. The \tp\ occurs at different times after the explosion for different models owing to the changes in \ek, \mej, and the mass and distribution of \Nifs. We choose times of -6, -2, 0, +3, +7, +12, and +21 days with respect to \tp. These times are chosen to cover $\sim 4$ weeks of observations, in order to capture the evolution of the main spectral features and cover a wide range of epochs from observed data.

\subsection{Helium Identification} \label{subsec:HE_ident}
In the following sections, we will identify \HeI\ lines by eye, rather than by a quantitative value. Measured values, such as equivalent width or optical depth, are model dependent and may not reflect observed conditions. Section \ref{subsec:abu_prof} showed that none of the models contain sodium.  The \NaI\,D line has a rest wavelength of 5895\,\AA\ and is both strong and common in SE--SNe. This line likely contributes to the total equivalent width of the 5875\,\AA\ region in observed events \citep{2018A&A...618A..37F}. For spectra with a strong 5875\,\AA\ line but weak or unobservable \HeI\,6678 and 7065 lines, the 5875\,\AA\ line may be dominated by Na but may have a weak contribution by helium, if traces of helium were present. As Na is not present in these models, the 5875\,\AA\ region has no \NaI\,D contribution.

Given these considerations, we describe the optical \HeI\ features in four ways; absent, weak, intermediate, and strong. Absent lines are those spectra which contain no \HeI\ lines as the He in the model is absent or the NLTE module is not used. Weak lines are difficult to easily identify in a spectrum. These lines may be blended with nearby features, broadened significantly, or simply not be a dominant feature in the region. In addition, the other two optical \HeI\ lines are absent or too weak to be separated from the spectral continuum. Intermediate strength lines are easier to identify, but may still show broadening or blending. The other two optical \HeI\ lines may also start to appear in the spectra, but may be weak. \HeI\ features identified as strong show signatures of all three optical \HeI\ lines. With regards to observational events or classification, weak or absent lines would be related to a pure Type Ic, intermediate would fit into transitional Type Ib/c or a Type Ic SN that shows some evidence of \HeI\, while strong lines would be a Type Ib.

\section{Light Curves}

\begin{figure}
    \centering
    \includegraphics[scale=.65]{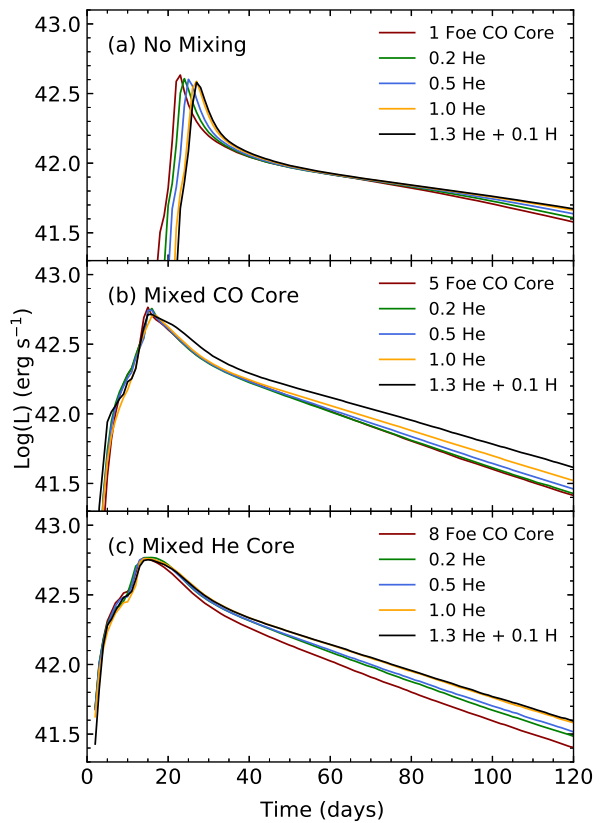}
    \caption{The radioactively powered light curves of the the 5 models at 1, 5, and 8 foes with the three mixing approximations.
    \label{fig:all_lcs}}
\end{figure}

Figure \ref{fig:all_lcs} shows light curves for 3 of the 4 explosion energies for all 5 models and mixing approximations. In the top sub-figure (Fig. \ref{fig:all_lcs}(a), the models with original composition), all 5 models have the an \ek\ of 1 foe and \Nifs\ mass of 0.096 \msun\ with the only difference being the total \mej. As the mass increases, the \tp\ occurs slightly later with \tp\,'s near 20-30 days after explosion. The extra mass traps more $\gamma$-rays $\sim$70 days after explosion producing brighter tails. The centralized \Nifs\ in these models prevents the early release of photons until the density lessens as the ejecta expands, resulting in a sudden release of photons and a rapid increase in luminosity.

The middle sub-figure (Fig. \ref{fig:all_lcs}(b)) shows the models with a mixed CO core at an \ek\ of 5 foes. The mixing approximation for \Nifs\ and \Fefs\ results in a very similar final \Nifs\ abundance pattern as seen in the third column of Figure \ref{fig:all_abund}. With a similar \Nifs\ abundance, \Nifs\ mass, and \ek, only the \mej\ changes and this affects the later light curve the most, showing more trapping in the model with 1.3 He + 0.1 H (black line) compared to the bare CO core model (red line).

The bottom sub-figure (Fig. \ref{fig:all_lcs}(c)) shows the models with a mixed He core at an \ek\ of 8 foes. The mass and distribution of \Nifs\ is the same for all levels of He stripping with the largest difference is the total \mej. The \tp\ occurs slightly earlier and the light curve is fairly broad. The ``bump" observed near 10 days is due to the expansion of the ejecta, lowering the innermost density to the point that the innermost trapped photons escape, resulting in the increase in luminosity towards the final peak.

\section{Bare CO Core Spectra} \label{sec:co_core}

In \citet{10.1093/mnras/staa123}, the model used to generate the Type Ic spectra was stripped of all He in the outer ejecta prior to collapse. For the CO core models in this work, there remains some amount of He left over from the stripping processes. Table \ref{tab:model_params} gives $\sim$0.02 \msun\ of He in the atmosphere of the models in this work and the \mej\ of these models are significantly more massive than those used in \citet{2012MNRAS.422...70H}. To check the validity of this limit for more massive and energetic SNe, the CO core spectra are generated with and without the non-thermal contributions of He. For the bare CO core spectra, we use the physical classification system from \citet{10.1093/mnras/stx980} with a slight modification. As the models contain no Na, we classify the spectra as Ic-\mN\,(\mN\,+1), where \mN\ is the average number of observed features in the spectra at and around \tp, and \mN\,+1 is the number of observed features plus the possible identification of \NaI\,D if Na was included in the abundances and if the line was observable. 

\subsection{Mixed Bare CO Core}\label{subsec:compl_cocore}

\begin{figure*}
    \centering
    \includegraphics[scale=.63,angle=0]{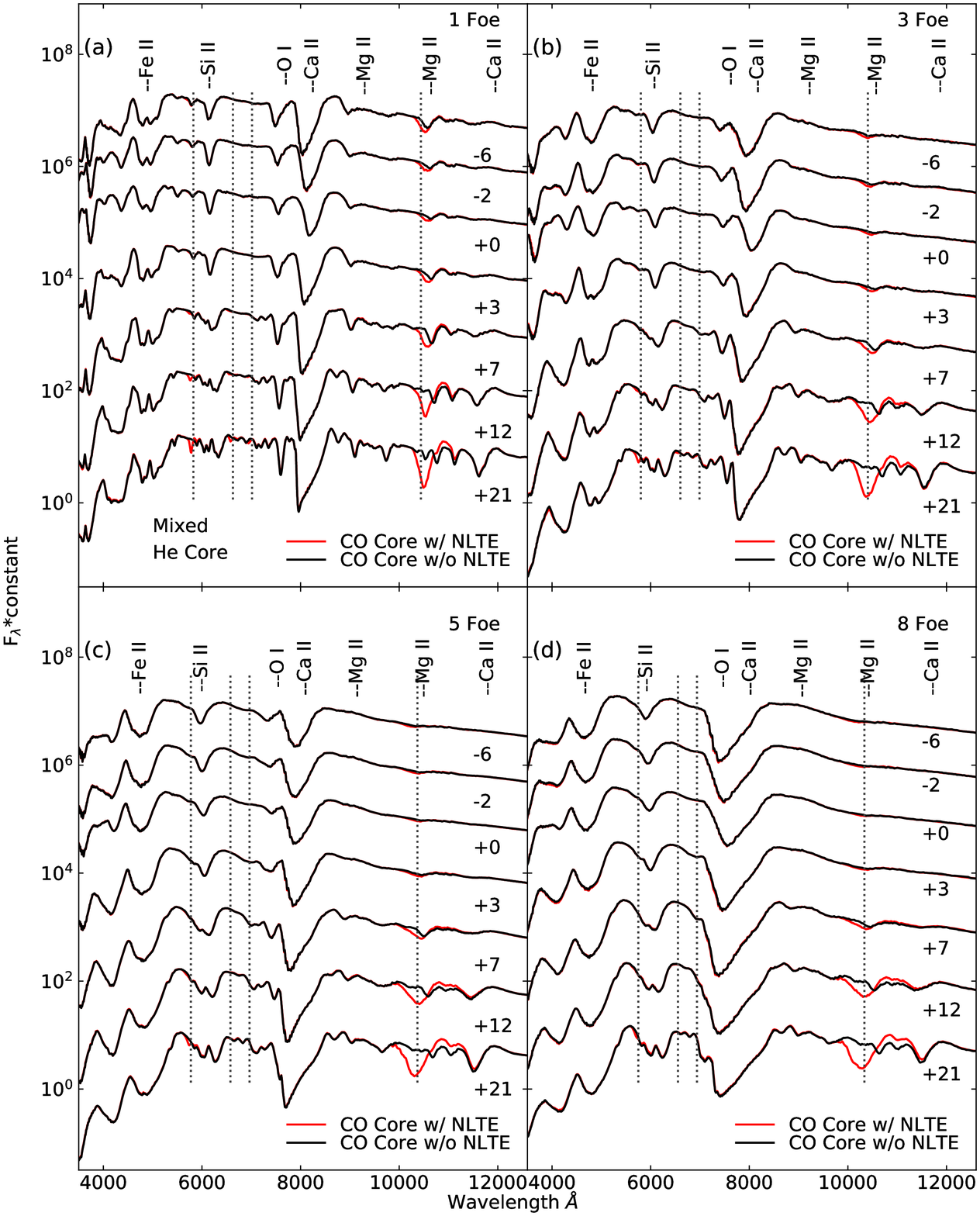}
    \caption{The mixed CO core models with the non-thermal contributions of He included in red and ignored in black. The y-axis is plotted on a logarithmic scale is arbitrarily scaled and shifted for visibility. The dashed grey lines represent the approximate positions of the 5875, 6678, 7065, and 10830 \AA\ \HeI\ lines. The identifiers at the top of each sub-figure are the approximate positions of these lines and may not be present in each spectrum.
    \label{fig:cocore_complete}}
\end{figure*} 

The optical 5875 \AA\, \HeI\, line in the 1 foe model (Fig. \ref{fig:cocore_complete}(a)) is narrow and weak, only becoming identifiable 12 days after \tp. As the photosphere recedes in the later epochs, the total \mhe\ available above the photosphere increases in the mixed CO models. If Na was present in this model, it is likely that the \NaI\,D line would be dominant in this region earlier than 12 days, making clear identification of the weak \HeI\ line difficult. As the energy increases in our models, the broadening of \HeI\, and nearby lines makes identification, even at 21 days after \tp, hard. At 8 foe, the difference between the two spectra at \tp\ + 21 days in the region near \HeI\,5875 is negligible.

The 1 foe model does show the presence of the two other optical \HeI\ lines at 6678 and 7065 \AA\, in the \tp\ + 21 days spectrum, but these two lines are extremely weak or absent in these models. Nearby to these two features are several \FeII\ lines that begin to increase in number and strength as $M_\mathrm{Fe}$ above the photosphere increases. The increased energy broadens the \FeII\ lines and effectively blends the \FeII\ and \HeI\ lines. The \HeI\,10830 line is weakly visible in the earliest epochs. This feature appears to make the stronger \MgII\ line appear broader instead, but does grow in strength as the ejecta evolves and more He is present to saturate the line. This contamination can make simple \HeI\ identification in observed SNe without spectral modelling challenging. As \ek\ increases, the mixed \mhe\ covers a wider range of velocities, producing broader \HeI, \MgII, and \CaII\ features in this wavelength range. 

The 1 foe bare CO core model in Fig. \ref{fig:cocore_complete}(a) would be classified as a Ic-5(6) or 6(7) based upon the \tp\ and nearby spectrum. In the three strongest \FeII\, lines near 5000 \AA, the bluer lines might be blended but no other features show blending due to the lower energy and velocity of the ejecta as the ejecta expands into the later epochs. The mixed CO core results in a somewhat homogenous spectral evolution in regards to location and observability of the primary SN\,Ic spectral features, with the main exception of narrow \FeII\ lines produced in the spectrum at 21 days after \tp\ due to the continued accumulation of Fe in the cooling ejecta. 

The 3 foe bare CO core model (Fig. \ref{fig:cocore_complete}(b)) would be classified as a Ic-5(6) based upon the \tp\ spectrum. Despite the appearance of a blended \FeII, the feature contains two notches that are still used to identify two separate \FeII\, features with other features being present and identifiable. When compared to the 1 foe models, the separation between the \CaII\, NIR triplet and \OI\ features has lessened due to the higher energy ejecta broadening the features. This extra broadening also affects the late--time narrow \FeII\ lines and \SiII\,6355 line as the spectra evolve.

The 5 foe bare CO core model (Fig. \ref{fig:cocore_complete}(c)) would be classified as an Ic-4(5) or a Ic-5(6) based upon the \tp\ spectrum and the near peak spectra. The double notch in the peak spectrum is difficult to discern but the \tp\ - 2 and + 3 day spectra show a slightly stronger signature of this double notch. The other lines important to the classification are not blended and easy to observe. The \OI\ feature is slightly broader, but still separate from the \CaII\, NIR triplet. The \FeII\ lines are still visible in the \tp\ + 21 days epoch and have continued to broaden. 

The 8 foe bare CO core model (Fig. \ref{fig:cocore_complete}(d)) would be classified as an Ic-3(4) based upon the \tp\ spectrum and the near peak spectra. The three strongest \FeII\ lines are now fully blended. The 7774 \AA\ \OI\ line and \CaII\, NIR triplet are also blended. The Type Ic given the Ic-3(4) classification in \citet{10.1093/mnras/stx980} are often broad lined Type Ic and can be associated with GRB's \citep{Mazzali_2001,10.1093/mnras/stz1588}. These events are often modelled with high \ek\ and significant mixing \citep{2002ApJ...572L..61M}.

\section{He-poor/rich Model Spectra} \label{sec:he_mods}

\citet{10.1093/mnras/staa123} showed that models with no mixing produced extremely blue spectra and light curves that did not match observed SNe. For this reason, we will not include such spectra in this work. 

In the following sections, we will first consider the \HeI\ lines observed in Type Ib SNe such as the 5875, 6678, 7065, 10830, and 20581 \AA\ \HeI\ lines and how the chosen mixing approximation, the \ek\ of the explosion, and the total mass of He affects the behaviour of these lines. He--poor will typically be used to label the models with 0.2 \msun\ of He while the label He--rich will refer to the models with 1.0 \msun\ of He. After the He analysis, we will discuss behaviour seen in the other features. \citet{10.1093/mnras/stx980} defined a classification system for Type Ib SNe. However, we cannot apply it here, because it focused on the weak, transient \Ha\ line that is not present in our Type Ib--like models.

\subsection{He-poor/rich Models with Mixed CO Cores}
\label{subsec:full_he_mods}
\begin{figure*}
    \centering
    \includegraphics[scale=.63,angle=0]{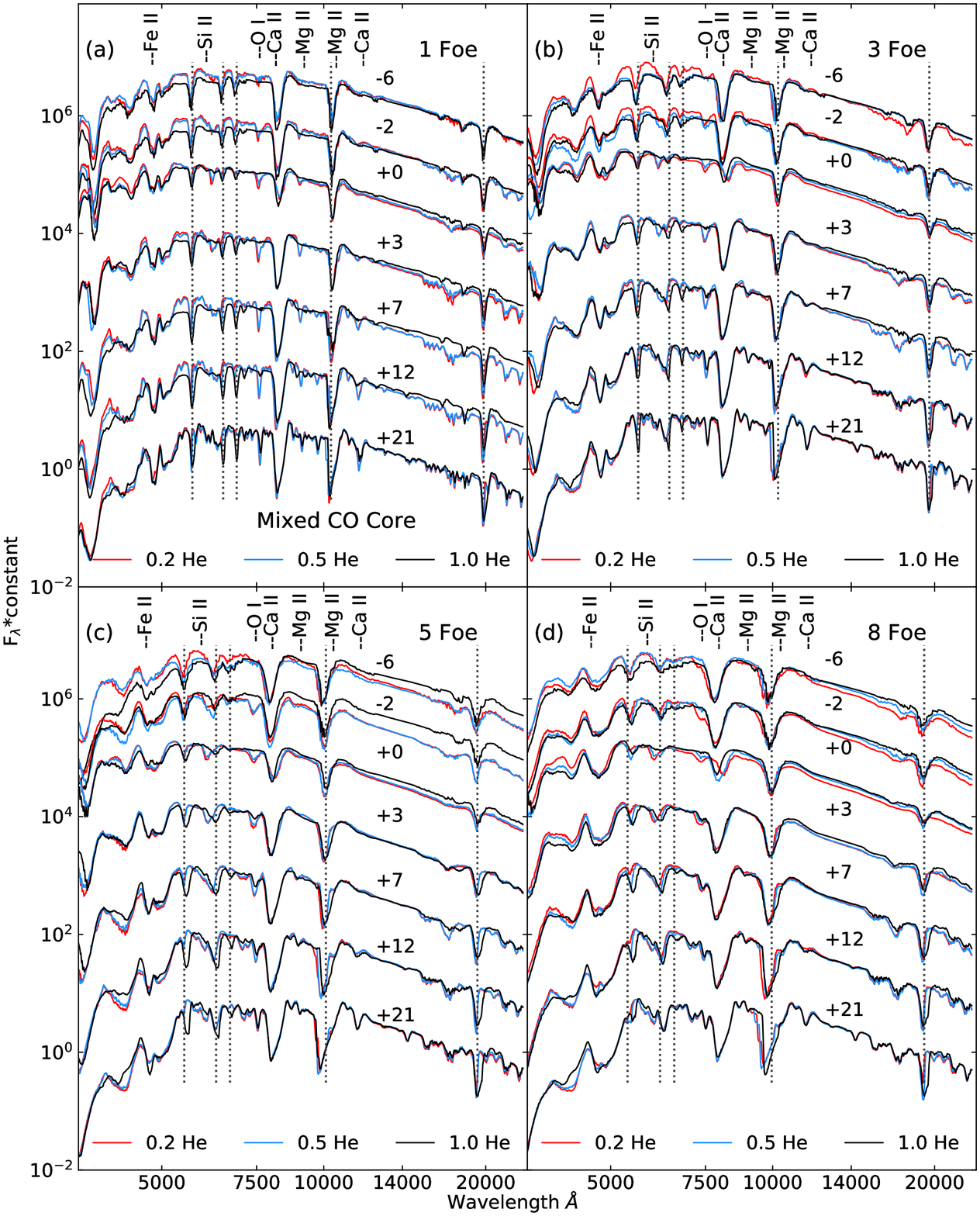}
   \caption{The spectral evolution of He-poor/rich models with mixed CO cores for all four \ek. The y-axis is plotted on a logarithmic scale is arbitrarily scaled and shifted for visibility. The 5 dashed grey lines represent the approximate positions of the 5875, 6678, 7065, 10830, and 20581 \angstrom\ \HeI\ lines. The identifiers at the top of each sub-figure are the approximate positions of these lines, which may not be present in every spectrum.
    \label{fig:allhe_full}} 
\end{figure*}

The 1 and 3 foe He-poor/rich models (Fig. \ref{fig:allhe_full}(a)-(b)) all show three narrow weak to intermediate optical \HeI\ lines in the pre-peak spectra. The 5875 \AA\ \HeI\ line is visible across all epochs for the models with 0.2 and 0.5 \msun\ of He, showing a weaker line at 12 and 21 days after \tp. In the 5 and 8 foe series (Fig. \ref{fig:allhe_full}(c)-(d)), the minima of the 5875 \HeI\ line in the He-poor models is shifted blueward compared to the He-rich model as the base of the line forming region is at a lower velocity. The 10830 \AA\ \HeI\ line shows little variation as a function of \mhe, \ek, or epoch, as this line typically saturates already with moderate excitation/ionization conditions for small He masses. As the \ek\ increases, this feature becomes slightly broader as the line core is effectively optically thick and only the line wings become stronger as the line forming region that contains He increases in velocity space. The 20581 \AA\ \HeI\ line is strong and identifiable at all epochs and explosion energies.

Fig. \ref{fig:allhe_full}(a) shows the 1 foe He-poor/rich models having narrow lines due to the low explosion energy of the models. The \OI\,7774 line is the most interesting feature at this energy as its evolution appears to be dependent on the \mhe. The model with 0.2 \msun\ of He shows a weak but visible \OI\ line for all epochs, while the model with 1.0 \msun\ of He only develops a strong \OI\ line 12 days after \tp. This has been observed for both Type Ib/IIb SNe \citep{10.1093/mnras/sty2719,2018A&A...618A..37F}, and is also seen in the \SiII\ line at 6100 \AA. The outer He rich shell contains negligible fractions of O/Si, so as the photosphere recedes into the mixed CO core, the O/Si abundance increases. As the \mhe\ increases, the size of the He shell also increases, resulting in the photosphere remaining in the He rich shell longer.

The increased energy of the 3 foe He-poor/rich models (Fig. \ref{fig:allhe_full}(b)) leads to a broadening of the observed lines, and the photosphere in the model with 1.0 \msun\ of He receding into the CO core earlier than in the 1 foe models. This results in the \OI\ line appearing strongly 7 days after \tp\, instead of 12 days. By 21 days after \tp, all three levels of He-stripping result in spectra with minimal differences. The relatively low energy of the explosions leads to little blending of strong features in the later spectra. The \OI\ and \SiII\ lines in the 5 foe He-poor/rich models (Fig. \ref{fig:allhe_full}(c)) are visible in the model with 1.0 \msun\ of He starting from 3 days after \tp. The increased energy of the ejecta broadens the \SiII\ and blends it into the \HeI\,6678 \AA\, line at later times. A secondary \SiII\ line near 5600 \angstrom\, has a lesser effect on the blue side of the 5875 \AA\ \HeI\ line. The \FeII\ lines show partial blending but the \CaII\ NIR triplet and \OI\ line remain split at all epochs.

The \OI\ and \SiII\ lines now become uniform across the 8 foe He-poor/rich models (Fig. \ref{fig:allhe_full}(d)) at approximately 7 days and 12 days respectively after \tp. Ignoring the He features, these spectra are nearly identical 7 days after \tp\, and onward, as the CO cores are similar in evolution and explosive nucleosynthesis, as seen in the first column of Fig. \ref{fig:all_abund}. As the mass of He above the CO increases, the outer edge of the CO core is shifted to a lower velocity which results in the photosphere entering the CO cores at earlier times for the He-poor model and later times for the He-rich model. Despite the high \ek, only the \FeII\ lines show blending.

\subsection{He-poor/rich Mixed He Core Model Spectra}\label{subsec:complete_he_mods} 
\begin{figure*}
    \centering
    \includegraphics[scale=.63,angle=0]{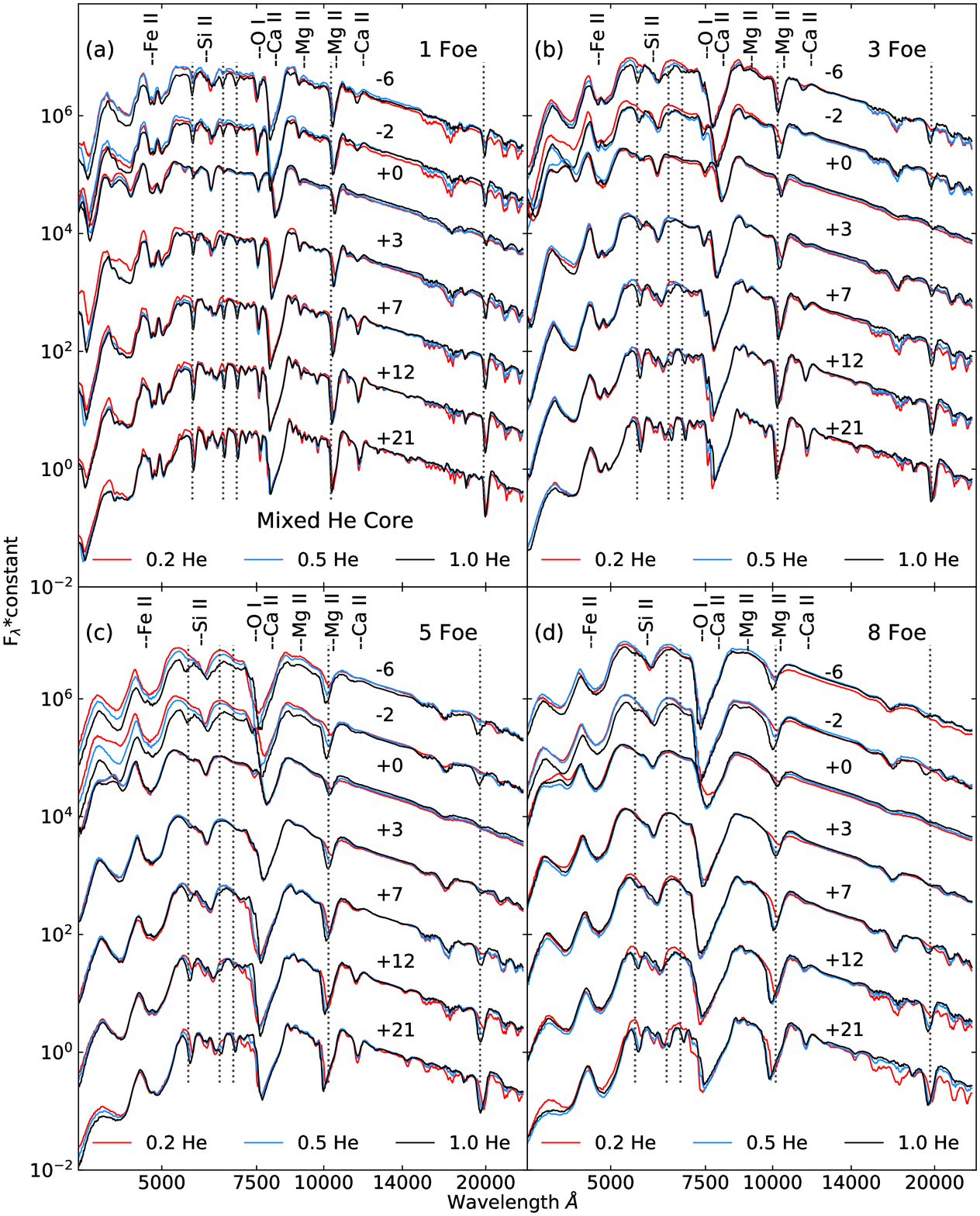}
   \caption{The spectral evolution of He-poor/rich models with a mixed He core for all four \ek. See Fig. \ref{fig:allhe_full} for plot description.
    \label{fig:allhe_comp}}
\end{figure*}

The He-poor/rich mixed He core models in Figure \ref{fig:allhe_comp} distribute \mhe\ across the entire ejecta. As \ek\ increases, the $\Delta$v in which \mhe\ is evenly distributed increases as the density decreases, meaning the average abundance per shell decreases. This requires that for the higher \ek\ models, the photosphere has to recede deeper into the ejecta to accumulate enough He to produce the optical \HeI\, lines.  The optical \HeI\ lines in the model with 1.0\,\msun\ of He, at 8 foe are intermediate and easily observable 7--12 days after \tp\ (Fig. \ref{fig:allhe_comp}(d)), while weak at 21 days after \tp\ for the model with 0.2\,\msun\ of He.

The \HeI\, 10830 line is strong for all models and all energies, and nearly all epochs. The 8 foe models show significant broadening of this line in the early epochs. This extra broadening and possibly weak signatures of P-Cygni profiles could result in harder He identification if one were to use this line for classification. Given that this wavelength range is less observed and often noisy, this would make clear identification of an He feature challenging. The \HeI\, 20581 line is strong at all energies but is limited near peak due to the higher luminosity and resulting temperature.
Similar to the mixed CO core models, the low \ek\ spectra (Fig. \ref{fig:allhe_comp}(a)) show narrow lines with minimal blending. The mixed He core results in similar line behaviours for all degrees of He stripping, excluding the \HeI\ lines themselves. Increasing the energy to 3 foe (Fig. \ref{fig:allhe_comp}(b)) produces broader lines with partial blending of certain features. The weak \HeI\ lines for the He--poor model at high energies in the early epochs could result in a Type Ic classification rather than Ib.

The 5 foe spectra in Fig. \ref{fig:allhe_comp}(c) show more line broadening and blending, such as the \OI\ and \CaII\ NIR triplet. The model with 1.0 \msun\ of He shows the least blending of these two features. Similar to the bare CO core at 8 foes, most features in Fig. \ref{fig:allhe_comp}(d) are broadened and blended for most or all epochs, such as the the \OI\ and the \CaII\ NIR triplet and the \FeII\ lines near 5000 \AA. The weak optical \HeI\ lines at this combination of \ek\ and mixing make the resulting spectra similar to the bare CO core at 8 foes in Fig. \ref{fig:cocore_complete}.

\begin{figure}
    \centering
    \includegraphics[scale=.50,angle=0]{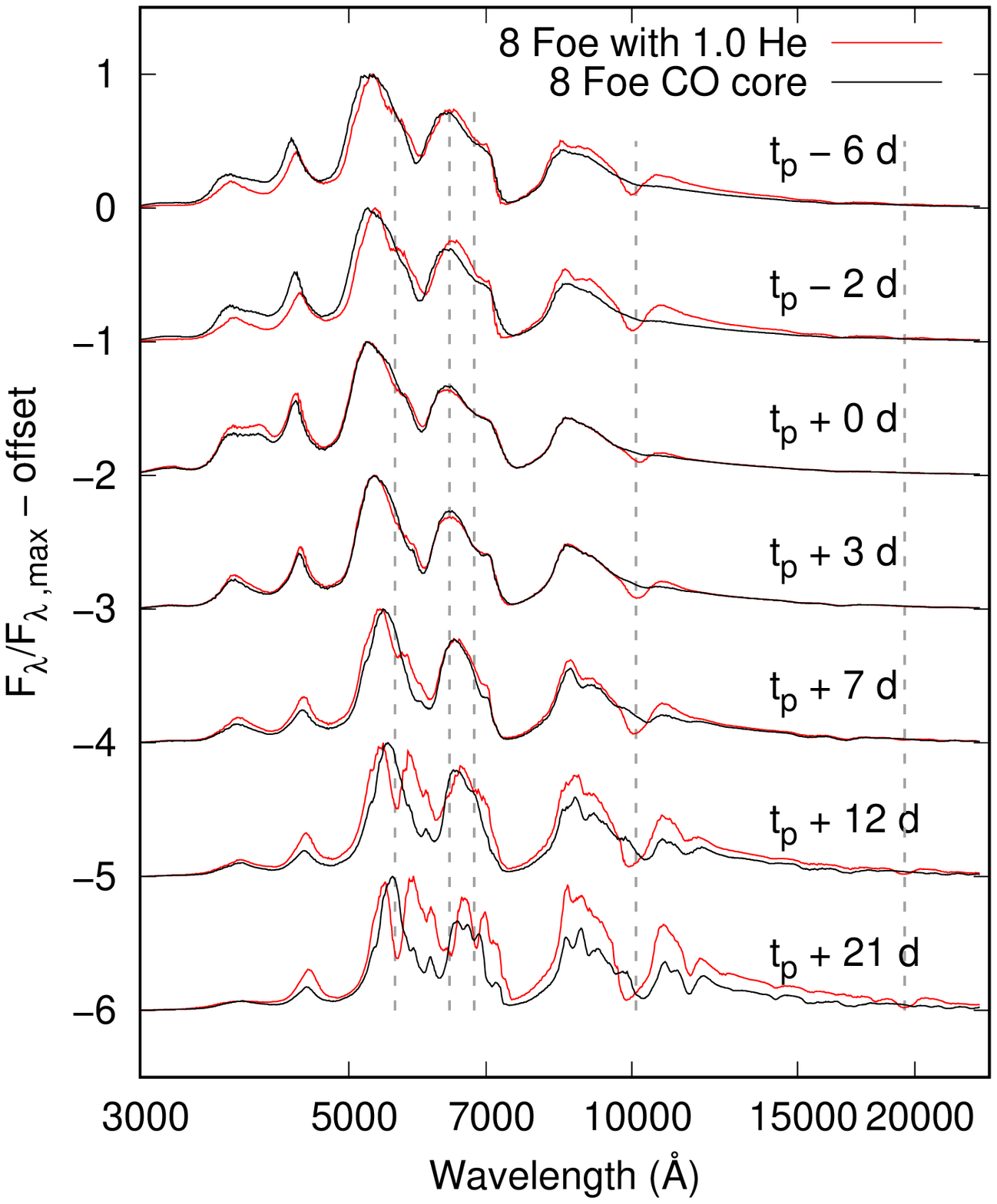}
    \caption{Mixed He Core model with 1.0 \msun\ of He compared to the He--free mixed bare CO core model exploded at 8 foes. The grey dashed lines are the approximate locations of the \HeI\ lines. \label{fig:cocore_8foe}}
\end{figure}

Figure \ref{fig:cocore_8foe} shows the mixed He core model with 1.0 \msun\ of He and the He--free mixed bare CO core model at \ek\ = 8 foes. Until 7 days after \tp, the only significant difference is the NIR 10830 \AA\ \HeI\ line. After this epoch, the differences in relative abundances and photospheric values result in different behaviour 12 and 21 days after \tp. The deeper features between 5500 and 6500 \AA\ of the bare CO core are a result of a deeper photosphere producing stronger \SiII\ and \FeII\ lines and a lack of re-emission by He. The bare CO core model does not produce a clean 10830 \AA\ \HeI\ line for any of the epochs. By day 7 to 12 after \tp, a combination of Mg, Ca, and O lines in this region, including weak \HeI\ lines, blend to form a feature similar to that of the He--dominated line in the model with 1.0 \msun\ of He. In addition, if Na was included in the models, the \NaI\,D line would likely be coincident with the strong optical \HeI\ line, further complicating identification of He if this was an observed SNe.


\section{H-poor/He-rich Model Spectra} \label{sec:h_mods}
The models with 1.3 \msun\ of He and 0.1 \msun\ of H are the last ones in the sequence of stripping that we consider in this work. The \Ha\ lines in these models are absorption dominated, so we will not consider the \R\ value used by \citet{10.1093/mnras/stx980} for these models. In the following sections, we will briefly discuss the optical \HeI\ and H lines. We include the H-free model with 1.0 \msun\ of He used in Section \ref{sec:he_mods} as a comparison.  After the H/He analysis, we then consider each spectral series as a function of energy and discuss other interesting features.

\subsection{H-poor/He-rich Models with Mixed CO Core}
\label{subsec:full_hy_mods}
\begin{figure*}
    \centering
    \includegraphics[scale=.63,angle=0]{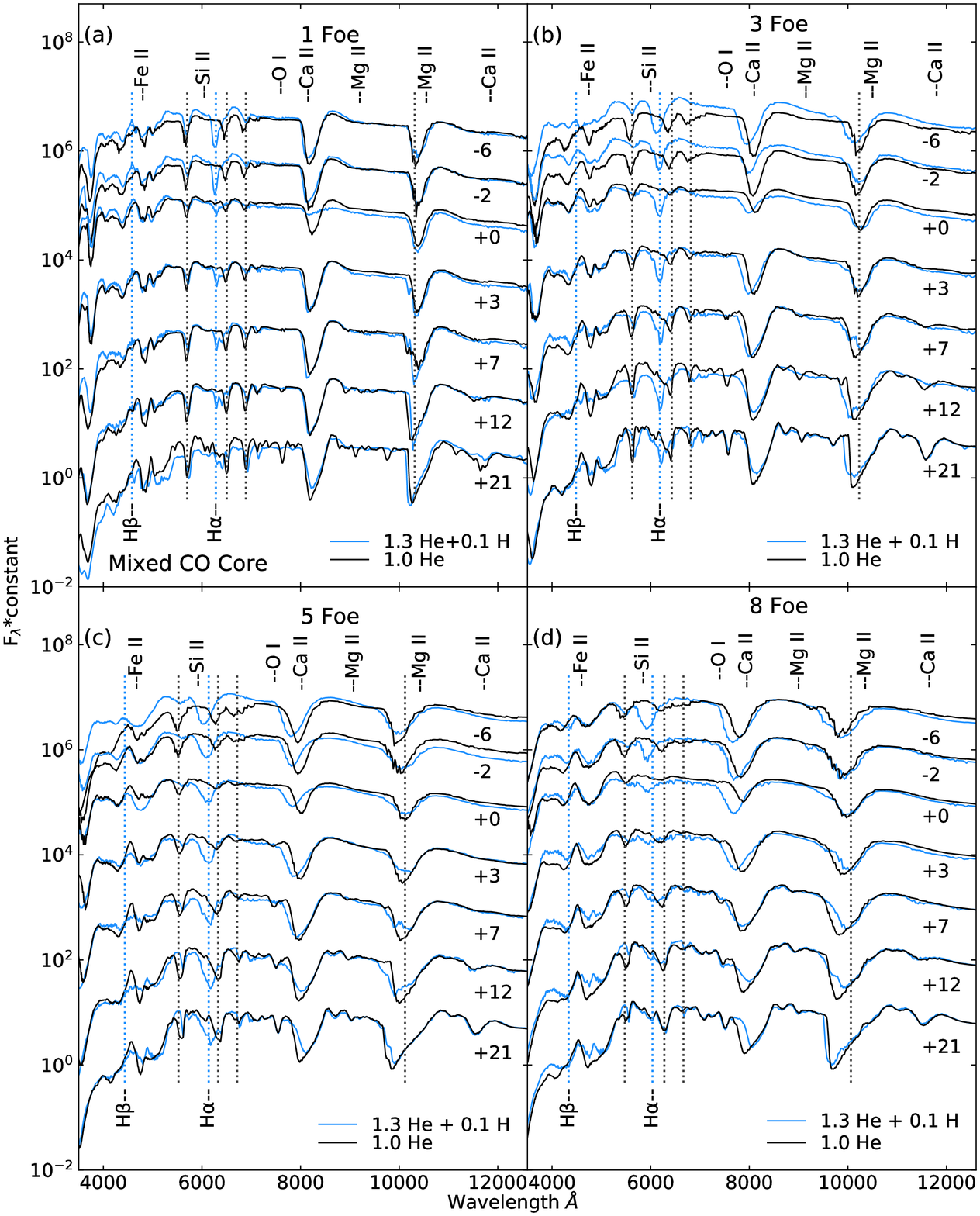}
    \caption{The spectral evolution of H-poor/He-rich models with a mixed CO core for all four \ek. The y-axis is plotted on a logarithmic scale is arbitrarily scaled and shifted for visibility. The four dashed grey lines represent the approximate positions of the 5875, 6678, 7065, and 10830 \angstrom\ \HeI\ lines while the dashed blue lines are the approximate positions of \Ha\ and \Hb. 
    \label{fig:hmods_full}}
\end{figure*}

The 1 foe spectra (Fig. \ref{fig:hmods_full}(a)) show a strong but narrow \Ha\ feature. The \Hb\ line is significantly weaker and hard to observe at this energy. The 3 foe explosion model (Fig. \ref{fig:hmods_full}(b)) shows a strong early \Ha\ line that maintains its strength for the entire evolution and is broader than the \Ha\ line in the 1 foe explosion model. The 5 and 8 foe explosion models (Fig. \ref{fig:hmods_full}(c-d)) continue the previous trends. The minimum of the \Ha\ line shifts redward as the spectra evolve, most noticeably in the 3, 5, and 8 foe models. The \Ha\ line starts to weaken and blend in the 8 foe case in the later epochs. At these epochs, the photosphere has receded into the CO core, and the multitude of weak lines blends out this region. The energy in the 1 foe model is too low to produce the same change in line velocity or this effect occurs in epochs earlier than the ones we consider.

The primary optical \HeI\ line at 5875 \AA\ is consistently observed in all models at slightly different minima. The re-emission from \Ha\ coincides with \HeI\,6678, causing this line to be difficult to observe directly in these models. This is most noticeable in the 1 and 3 foe models. The blending from the high velocity ejecta makes these lines hard to separate in the higher energy models.

The 1 foe H-poor/He-rich model (Fig. \ref{fig:hmods_full}(a)) has the photosphere and resulting line forming region remain predominately in the H/He rich shells outside the fully mixed CO core. This is most identifiable in the behaviour of \OI\,7774, which is weak for the 7 epochs shown in the model with 1.3\,\msun\ of He + 0.1\,\msun\ of H. The 3 foe H-poor/He-rich model (Fig. \ref{fig:hmods_full}(b)) follows a similar evolution but with broader lines. At 21 days after \tp, the spectra shows \SiII\ and \OI\ lines in the spectra. 

The 5 foe H-poor/He-rich model (Fig. \ref{fig:hmods_full}(c)) shows the \FeII\ lines being initially fully blended but become split $\sim$7 days after \tp. The other observed lines behave similarly to the 3 foe He-rich model, as the predominant changes are related to the \HeI\ or \Ha\ lines or slight changes in photospheric properties. The 8 foe H-poor/He-rich model (Fig. \ref{fig:hmods_full}(d)) still shows mostly unblended features. The photosphere recedes into the mixed CO core between 7 and 12 days after peak, producing a 7774 \AA\ \OI\ line similar to that of the H-poor/He-rich model at 5 foes. 

\subsection{H--poor/He--rich Mixed He Core Model Spectra}\label{subsec:complete_hy_mods}
\begin{figure*}
    \centering
    \includegraphics[scale=.63,angle=0]{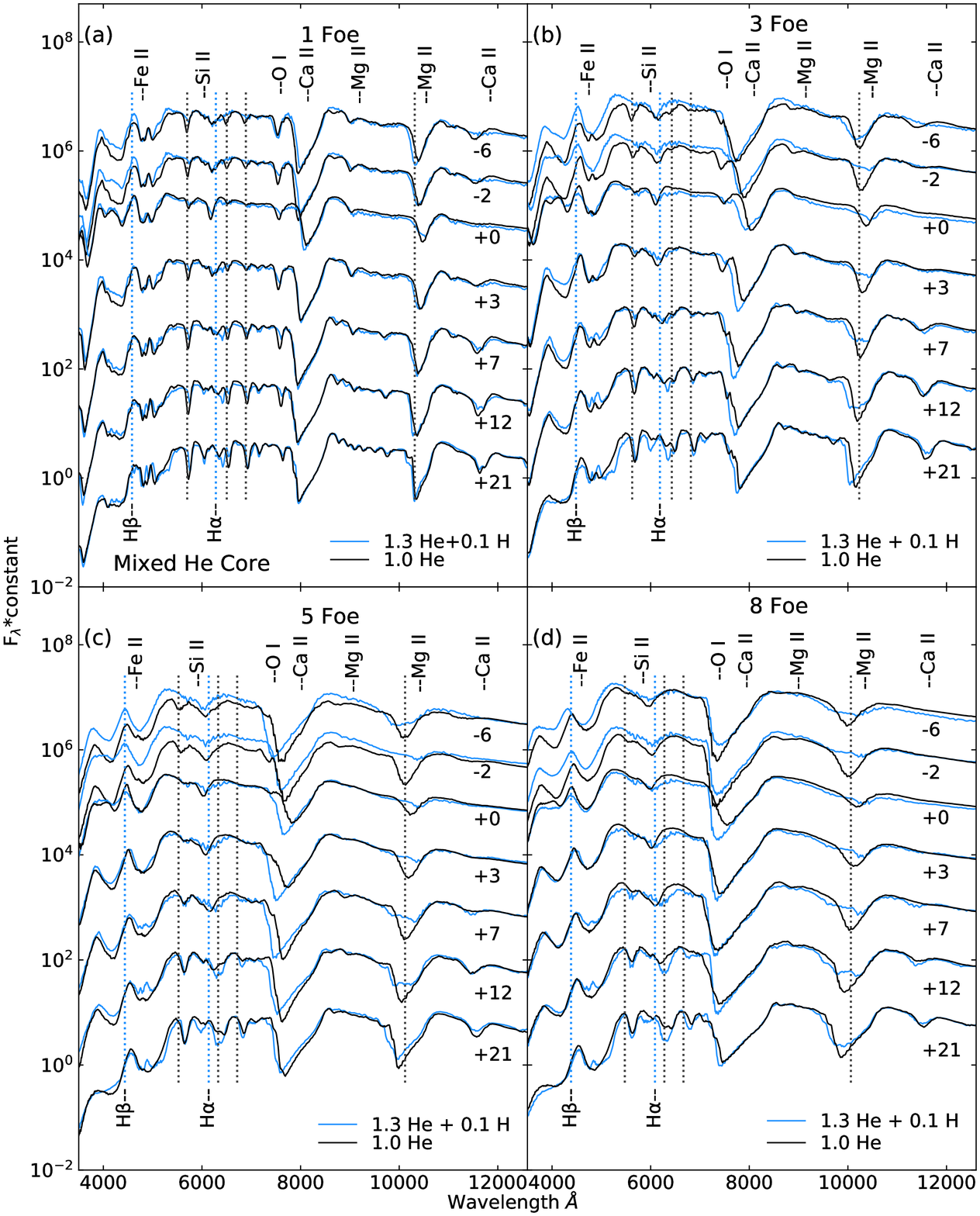}
    \caption{The spectral evolution of H-poor/He-rich models with a mixed He core for all four \ek. See Figure \ref{fig:hmods_full} for plot description.
    \label{fig:hmods_complete}}
\end{figure*}

These models (Fig. \ref{fig:hmods_complete}) have 0.1 \msun\ of H evenly distributed throughout the 5.5 to 6 \msun\ ejecta with an average abundance of 1.6 - 1.8\%  across the entire model. As the \ek\ increases from 1 to 8 foes, the ejecta expands rapidly and the initial photospheric position for the earliest epoch is deeper into the ejecta. The deeper photosphere has more H mass in the line forming region, but is spread over a larger $\Delta$v as the \ek\ is increased, leading to significant broadening. These two factors result in the \Ha\ line only being easily identifiable approximately 7 days after \tp\, for all 4 \ek\, with the 5 and 8 foe models having the stronger and broader lines. The \Hb\ line is partially observable in the 1 and 3 foe models, but the broadened \FeII\ lines blend in this region, making easy identification in the 5 and 8 foe models challenging. The optical \HeI\ lines only show slight differences compared to the model with 1.0 \msun\ of He for all \ek. The two models shown in Fig. \ref{fig:hmods_complete} show some differences due to the significant amount of He in per zone in the mixed He cores. A completely mixed He core model with H is unlikely to be observed, given the velocity separation between the rich CO core and the He shell shown in the first column of Figure \ref{fig:all_abund}.

\section{Observed Type Ib SNe}\label{sec:obs_Ib_sne}

\begin{figure}
    \centering
    \includegraphics[scale=.55,angle=-90]{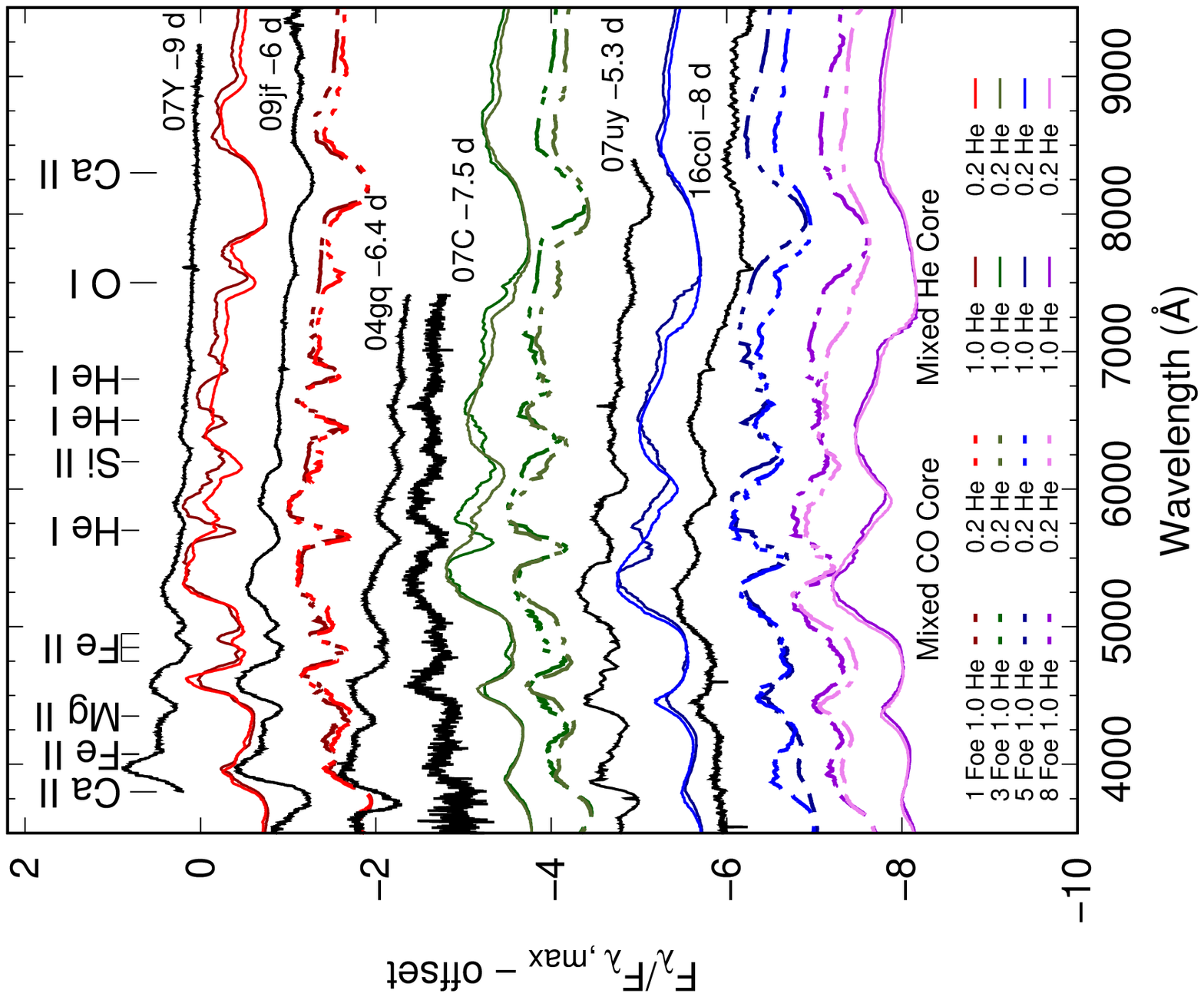}
    \caption{The 0.2 and 1.0 \msun\ of He models for all \ek\ and both mixing approximations at \tp\ - 6 days compared to a small sample of Type Ib SNe and the Type Ic SN SN2016coi.
      \label{fig:early_epoch}}
\end{figure}
\begin{figure}
    \centering
    \includegraphics[scale=.55,angle=-90]{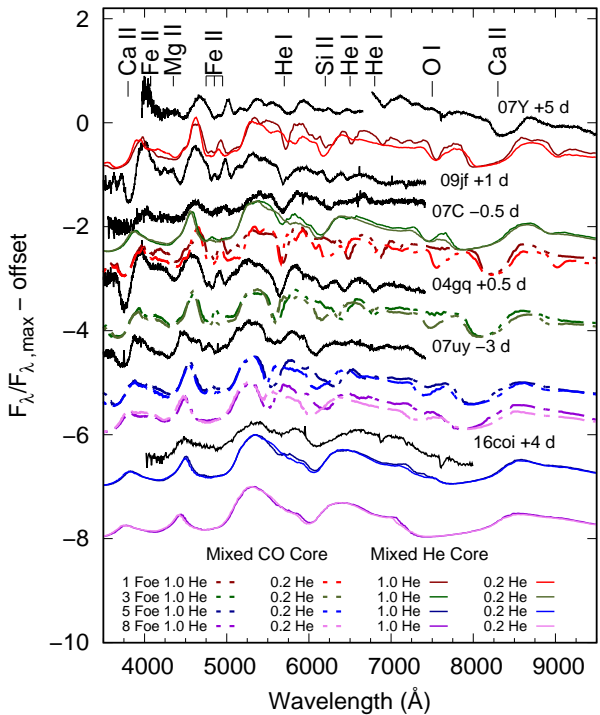}
    \caption{The 0.2 and 1.0 \msun\ of He models for all \ek\ and both mixing approximations at \tp\ + 3 days. The 5 and 8 foe mixed He Core model spectra do not show a strong \HeI\ line and are placed at the bottom.
    \label{fig:nearpeak_epoch}}
\end{figure}
\begin{figure}
    \centering
    \includegraphics[scale=.55,angle=0]{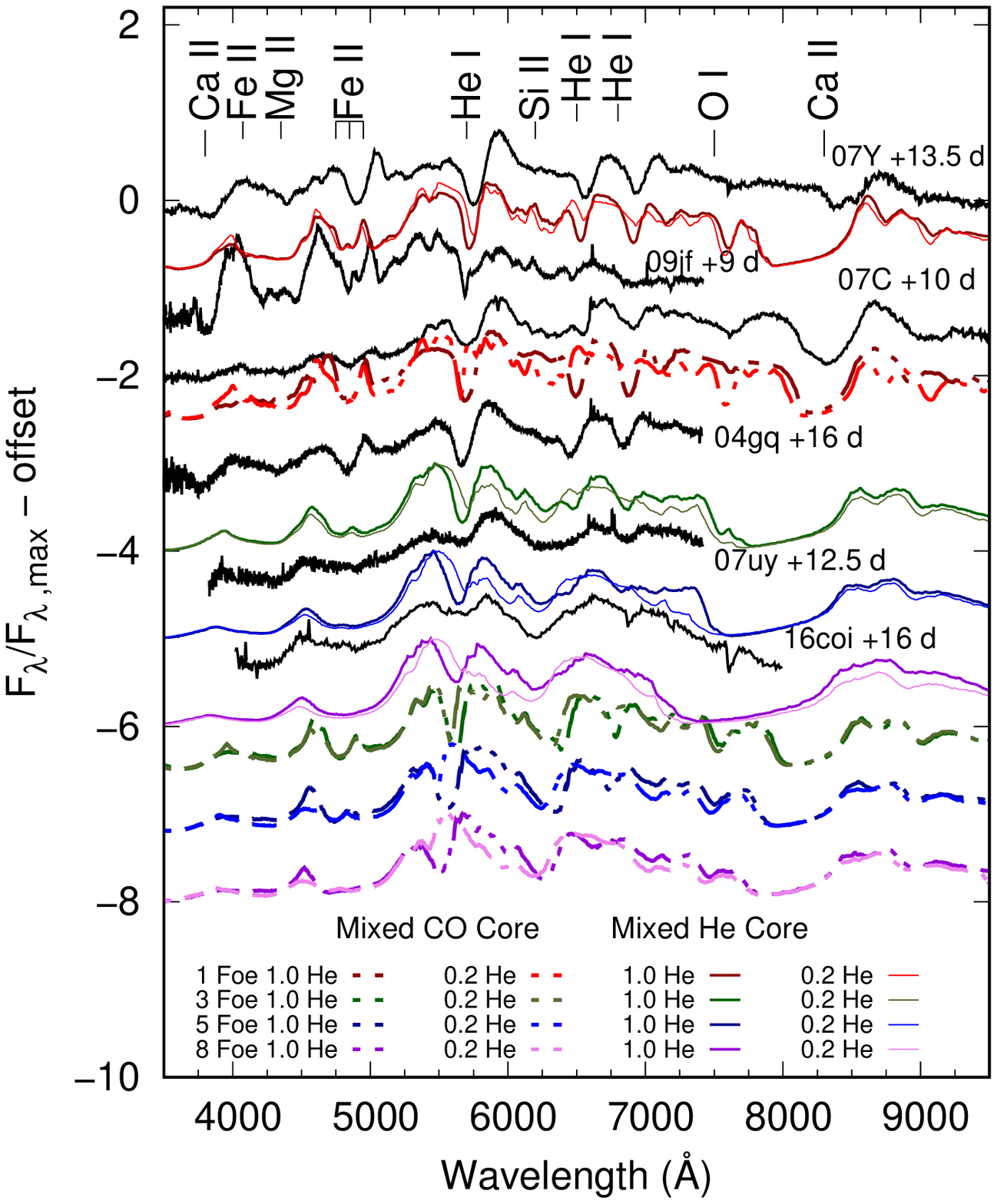}
    \caption{The 0.2 and 1.0 \msun\ of He models for all \ek\ and both mixing approximations at \tp\ +12 days.
    \label{fig:later_epoch}}
\end{figure}

In this section, we compare the synthetic spectra from the He-rich models for both mixing approximations and all four explosion energies to that of a small sample of Type Ib SNe at 3 epochs of approximately 1--2 weeks prior to \tp, $\pm$5 days near \tp, and $\sim$2 weeks after \tp. The SNe used in the following sections are chosen due to having a well sampled spectral series across a similar time range with respect to \tp\ that is used in our synthetic spectral models. In addition, SNe with explosion parameters (\ek\ and \mej\,) found in the literature that are near to the values in Table \ref{tab:model_params} are preferred. All \tp\, values used for the observed SNe in the following sections are found using the maximum luminosity measured with respect to a pseudo-bolometric light curve from \citet{2018MNRAS.478.4162P} for SN2016coi and \citet{doi:10.1093/mnras/stw299} for all others considered. All data in the following sections has been taken from WISeREP\footnote{https://wiserep.weizmann.ac.il/} \citep{2012PASP..124..668Y}.

In the previous sections, we generated a large set of synthetic spectra with varying He masses. We will compare these SNe to the models with 0.2 and 1.0\,\msun\ of He for all energies and both mixing approximations. The spectral comparisons shown in the following sections are scaled relative to each individual spectrum's maximum flux value. Due to the scaling and lack of fine tuned parameters, the ``best fit'' to an observed SN by one of the synthetic spectra is based upon a qualitative look at the spectra, and not a modelled ``best fit''. This ``best fit'' may not match all or most of the features, the spectral colour, or the epoch, but is the closest synthetic spectrum in the models with 1.0\,\msun\ of He. For these Type Ib SNe, if the 5875 \HeI\ is visible, the  \citet{10.1093/mnras/staa123} used a similar method to estimate the energy of a sample of SNe\,Ic, which showed promise as a basic approximation.

\subsection{SN\,2007Y} \label{subsec:sn07Y}

SN\,2007Y is a He-rich SN\,Ib with wide wavelength photometric coverage from radio to X-ray and is thought to be a very low energy SN with \ek\, $\approx 0.1$\,foe and a 3.3\,\msun\ He core. To date, only a nebular model has been produced \citep{2009ApJ...696..713S}. The nebular model estimates the energy and the mass to be significantly lower than the models in this work. The most similar synthetic spectra in Figures \ref{fig:early_epoch}, \ref{fig:nearpeak_epoch}, and \ref{fig:later_epoch} is the 1 foe mixed He core model with 0.2\,\msun\ of He. This model experiences significant amounts of mixing but the low \eom\ ratio results in narrow lined features at low line velocities.  The two earlier spectra show a far bluer shape than in the synthetic spectra. In the later epochs, SN2007Y has redder features with lines at a consistently lower velocity, suggesting that the event is less energetic than the comparable 1 foe model, but still shows a fairly similar overall spectral appearance.

\subsection{SN\,2007uy} \label{subsec:sn07uy}

SN\,2007uy is a He-rich SN\,Ib with a more comparable estimated \mej\, $\sim 4.4$\,\msun\ and an estimated \ek\,$\sim 15$\,foe \citep{2013MNRAS.434.2032R}, but no detailed model has been produced for this event. In Fig. \ref{fig:early_epoch}, the early spectrum of SN2007uy is best placed somewhere between the 3 foe mixed CO core model and the 5 foe mixed He core model. The \FeII\ features are not completely blended at this epoch and the redder \CaII\ NIR triplet is much closer to the 3 foe models. In Fig. \ref{fig:nearpeak_epoch}, the near peak spectrum is now much closer to the 3--5 foe models with a fully mixed CO core. Based on this, an estimated energy from the matched models is only a half to a fifth of the estimated \ek\ from \citet{2013MNRAS.434.2032R}. In the final spectrum shown in Fig. \ref{fig:later_epoch}, the \FeII\ features appear blended and relatively broadened, similar to the 3 and 5 foe mixed He core models but with a weaker \HeI\ line. \citet{2013MNRAS.434.2032R} discusses the aspheric evolution of several line forming regions, but do not find asphericity in the distribution of the \Nifs\, which would explain the similarities of both the two mixing approximations while maintaining a similar \ek.

\subsection{SN\,2009jf} \label{subsec:sn09jf}

SN\,2009jf is a He-rich Type Ib SN with no detailed model and only qualitative estimates for the \mej\ and \ek, given as 4--9 \msun\ and 3--8 foes in \citet{2011MNRAS.413.2583S}. \citet{2011MNRAS.416.3138V} use Arnett methods (\citet{1982ApJ...253..785A}) and estimate the mass to be 5--7 \msun\ with an \ek\ of 5.5--10.5 foe, but note that they suspect the energy estimate to be too high. In Fig. \ref{fig:early_epoch}, the strong \FeII\ lines and narrow features places the observed spectrum between 1 foe mixed He core model and 1 foe mixed CO core model. If we neglect the \OI\,/\CaII\ NIR blending, then SN2009jf compares slightly better to the 3 foe mixed He core model. However, as the SN evolves towards peak and 2 weeks after peak, the features continue to remain narrow and fits best with the 1 foe mixed He core model. \citet{2011MNRAS.413.2583S} suggests that the early appearance of \HeI\ lines at \tp\ - 14 days may imply that the ejecta is reasonably mixed. The visibility of what might be \SiII\,6355 and the \OI\,7774 in the early epoch may also suggest that the ejecta are either mixed well or the He shell is thin.

\subsection{SN\,2016coi} \label{subsec:sn16coi}

SN\,2016coi is a well studied and observed SN\,Ic-BL/Ic-4 that may have had a thin shell of residual helium left over prior to collapse \citep{2018MNRAS.478.4162P}. To date, no SN\,Ib-BL have been conclusively observed or been given a consistent classification. While the presence of He in SN\,2016coi is relatively uncertain, we include this event here to show a possible case of a broad-lined high energy event with He. All three spectra fit between the 5 and 8 foe models at both mixing approximations. At early epochs, due to the mixing approximation, much less than the total 1.0\,\msun\ of He is available in the line forming region above the photosphere, resulting in weaker early \HeI\ lines.

\subsection{SN\,2007C} \label{subsec:2007c}

SN2007C is a SN\,Ib with no modelling or detailed studies available in the literature and has only been included in bulk studies \citep{2011ApJ...741...97D,2014AJ....147...99M,doi:10.1093/mnras/stw299}. As seen in in Fig. \ref{fig:early_epoch} and \ref{fig:nearpeak_epoch}, SN2007C fits closest to the 3 foe mixed He core model showing fairly broad features and blending, particularly in the 4000--5000 \AA\ region. In Fig. \ref{fig:later_epoch}, the 4000--5000 \AA\ region is largely blanketed while the rest of the spectrum appears to be more narrow and well defined, matching that of the 1 foe model with a mixed CO core. This may suggest that the outer atmosphere is mixed well, including Fe group elements, or has a thin He shell before transitioning to a strongly mixed CO core.

\subsection{SN\,2004gq}\label{subsec:2004gq}

Similar to SN\,2007C, SN\,2004gq is a SN\,Ib with few references in the literature beyond bulk studies. In Fig. \ref{fig:early_epoch}, it is best compared to the 3 foe mixed He core model. By the near-peak epoch in Fig. \ref{fig:nearpeak_epoch}, SN\,2004gq has well defined \FeII\ lines, is still blue in colour, and looks like the 1 and 3 foe models with a mixed CO core. In the final epoch shown in Fig. \ref{fig:later_epoch}, SN\,2004gq is similar to the 1 foe model with a mixed CO core but also to the 3 foe He core model; the 4000--5000\,\AA\ region remains more similar to that of the lower energy 1 foe models for both mixing approximations.

\section{Observed SNe IIb}
\label{sec:obs_IIb_sne}

\begin{figure}
    \centering 
    \includegraphics[scale=.55,angle=-90]{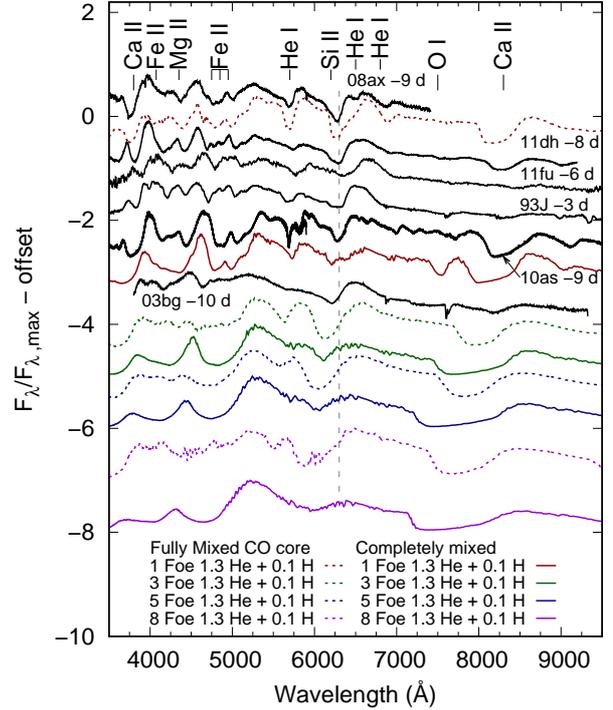}
    \caption{The 1.3 \msun\ of He + 0.1 \msun\ of H models for all \ek\ and mixing approximations at \tp\ - 6 days, compared to a small sample of Type IIb SNe. The dashed grey line represents the approximate \Ha\ line location based upon the 1 foe models and may not track the \Ha\, line in the higher energy SNe.
      \label{fig:IIb_early_epoch}}
\end{figure}

\begin{figure}
    \centering
    \includegraphics[scale=.55,angle=-90]{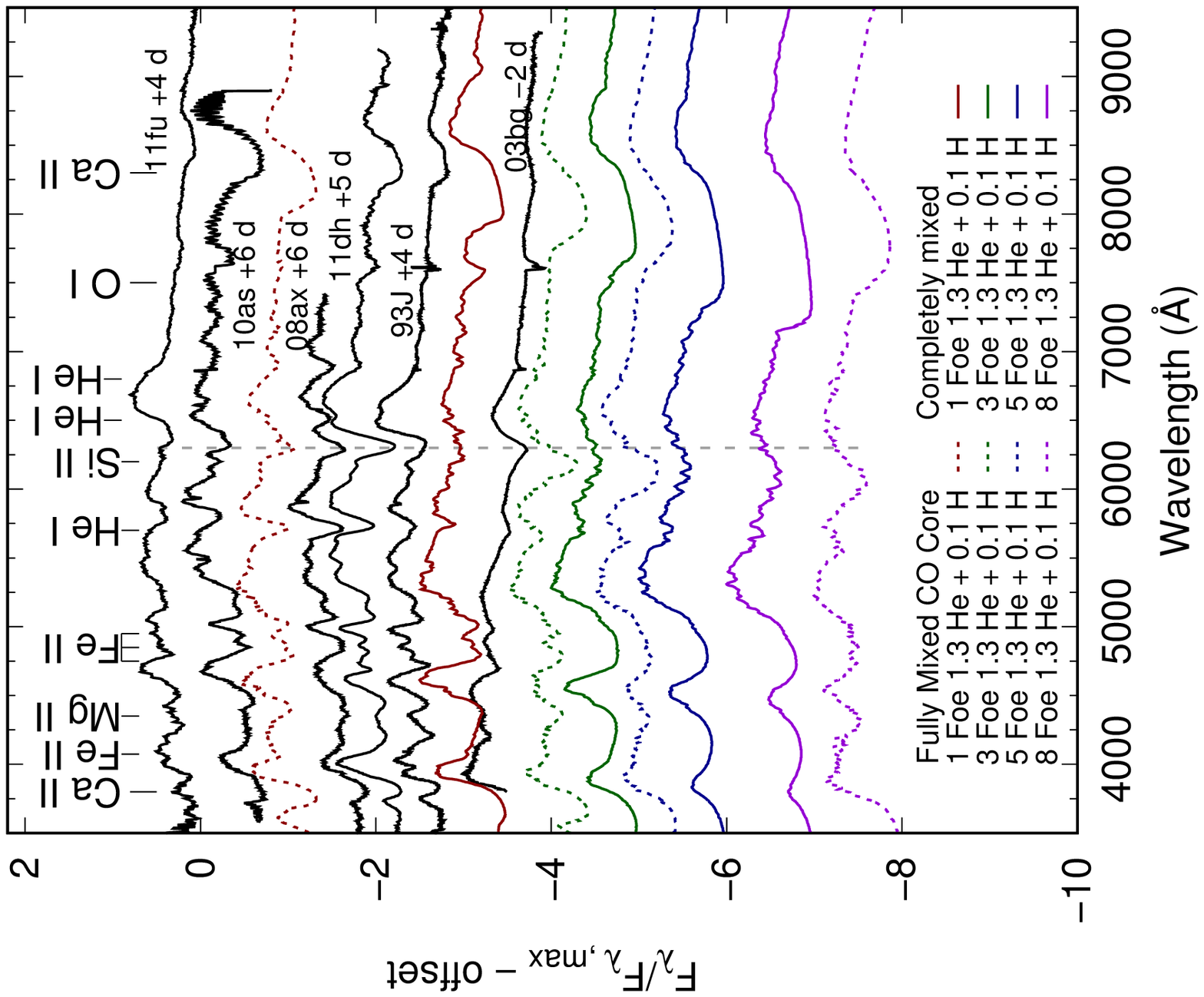}
    \caption{The 1.3 \msun\ of He + 0.1 \msun\ of H models for all \ek\ and mixing approximations at \tp\ + 3 days, compared to a small sample of Type IIb SNe. See Fig. \ref{fig:IIb_early_epoch} for plot explanation.
    \label{fig:IIb_nearpeak_epoch}}
\end{figure}

\begin{figure}
    \centering 
    \includegraphics[scale=.55,angle=-90]{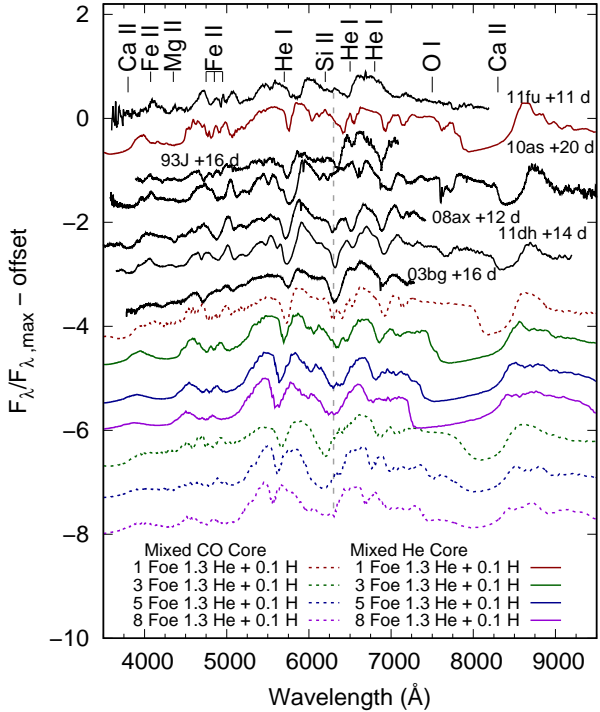}
    \caption{The 1.3 \msun\ of He + 0.1 \msun\ of H models for all \ek\ and mixing approximations at \tp\ +12 days, compared to a small sample of Type IIb SNe. See Fig. \ref{fig:IIb_early_epoch} for plot explanation.
    \label{fig:IIb_later_epoch}}
\end{figure}

We now consider the Type ``IIb-like" synthetic spectra from the H-poor/He-rich models with 1.3\,\msun\ of He + 0.1\,\msun\ of H, for both mixing approximations and all four explosion energies. We compare these to a small sample of SNe\,IIb under the same selection criteria for the SNe\,Ib in Section \ref{sec:obs_Ib_sne} at the same epochs.

\subsection{SN\,1993J}\label{subsec:SN1993J}

SN\,1993J is one of the first discovered SN\,IIb and is estimated to have 2--5 \msun\ of ejecta, with a thin shell of H remaining after binary companion stripping, and an estimated \ek\ of $\sim$1 foe \citep{1993Natur.364..507N,1993Natur.365..232S,1995ApJ...449L..51Y}. Comparing SN1993J to our Type IIb--like models in Figs \ref{fig:IIb_early_epoch} and \ref{fig:IIb_nearpeak_epoch}, the best fit location is between the 1 foe model with a mixed CO core and the 1 foe mixed He core model, but does not show strong similarity with either. The final spectrum in Fig. \ref{fig:IIb_later_epoch} shows similarity to those of the 1 foe models with both mixing approximations.

\subsection{SN\,2003bg}
\label{subsec:SN2003bg}

SN\,2003bg is a broad-lined SN\,IIb with a modelled \mej\,$\sim 4$\,\msun\ and \ek\,$\sim 5$\,foe \citep{2009ApJ...703.1624M,2009ApJ...703.1612H}. The \mej\ is slightly lower than the models in this work while \ek\ falls within our range.  Despite the similarities in model properties, the early spectrum in Fig. \ref{fig:IIb_early_epoch} does not show a good match to any of the synthetic spectra. The near peak spectrum in Fig. \ref{fig:IIb_nearpeak_epoch} shows some similarities to those of the 3 foe models with similar line velocities, but is still too blue and has a low relative flux. The final epoch in Fig. \ref{fig:IIb_later_epoch} is more comparable to the 1 foe mixed He core model.

\subsection{SN\,2008ax}
\label{subsec:SN2008ax}

SN\,2008ax is a well studied SN\,IIb with an estimated \mej\ of 2--5 \msun\ and an \ek\ of at least 1 foe \citep{2011MNRAS.413.2140T,2011ApJ...739...41C}. \cite{2011ApJ...739...41C} find spectropolarimetric measurements that give both early and late time deviations from axisymmetry. When placing SN\,2008ax in Figs. \ref{fig:IIb_early_epoch} to \ref{fig:IIb_later_epoch}, the spectra are similar to that of the 1 foe model with a mixed CO core, which show little to no blending. Of the Type IIb SNe we consider, SN\,2008ax and the 1 foe model with a mixed CO core show the most similar spectral behaviour.

\subsection{SN\,2010as}
\label{subsec:SN2010as}

SN\,2010as is another SN\,IIb, initially classified as a transitional SN\,Ib/c before becoming an unusual SN\,Ib with an estimated \mej\,$\sim$\,2--4 \msun\ and \ek\,$\sim$\,0.7 foe \citep{Folatelli_2014}. Similar to the previous SNe\,IIb, the best fit for SN\,2010as is somewhere between the mixed He core and mixed CO core models exploded at an \ek\,$= 1$ foe. The earliest spectrum (Fig. \ref{fig:IIb_early_epoch}) shows a strong similarity to that of the 1 foe mixed He core model at a lower velocity, before the lines began to narrow and then has more in common with the 1 foe mixed CO core model. This may suggest the outermost atmosphere, responsible for the early epochs, is more mixed than the deeper layers.

\subsection{SN\,2011dh}
\label{subsec:SN2011dh}

SN\,2011dh is another well-studied SN\,IIb with an estimated \mej\,$\sim 2$\,\msun\ and an \ek\ of 0.6--1.0 foe \citep{2011MNRAS.411.2726B,2012ApJ...757...31B,ergon2015}. The low \mej\ and \ek\ place it lower than any of the models in this work. In Figs. \ref{fig:IIb_early_epoch} to \ref{fig:IIb_later_epoch}, SN\,2011dh is most similar to the 1 foe models of both mixing approximations. The earliest spectrum shows what is likely to be a \CaII\, feature at 8500\,\AA, similar to the strong early signature of \CaII\ NIR triplet in SN2010as. These SNe are estimated to have a similar \ek\,/\mej\, ratio and share some commonalities in the remaining spectral evolution.

\subsection{SN\,2011fu}
\label{subsec:SN2011fu}

SN\,2011fu is a SN\,IIb that shares a similar doubly peaked light curve with SN\,1993J and using hydrodynamic modelling, has an estimated \mej\,$\sim 3.5$\, \msun\ with an \ek\,$\sim 1.3$\,foe \citep{10.1093/mnras/stv1972}. It shares a similar spectral evolution as SN\,1993J and the other SNe that fit between the 1 foe mixed He and Co Type IIb-like models in Figs. \ref{fig:IIb_early_epoch} to \ref{fig:IIb_later_epoch} but with line velocities shifted redward.

\section{Discussion} \label{sec:disc}

\subsection{Effects and Signatures of Mixing and Energy}
\label{subsec:class_bareco}

Observed H/He lines in Type Ib/IIb SNe typically show higher velocities pre-peak before a decline into slower and flatter velocities \citep{10.1093/mnras/sty3399,2018A&A...618A..37F}. This is thought to be related to the depth of the H/He line forming region and gives evidence of the abundance structure of the ejecta. This is directly seen in the behaviour of the \Ha\ line in the models in Figure \ref{fig:hmods_full}(c--d). As the photosphere recedes deeper into the ejecta as the spectra evolve, the approximate line velocity of \Ha\ shifts redward before plateauing as it reaches the end of the H-rich material. This is harder to observe in the He-rich models but may be more visible if earlier epochs were generated.

For the bare CO cores in Fig. \ref{fig:cocore_complete}, the 8 foe model would likely be classified as a Type Ic-3(4)/BL due to the extensive line blending in the spectra.  For the He-poor/rich mixed CO core models at an \ek\ of 8 foes (Fig. \ref{fig:allhe_full}(d)), the spectra show a split between the 7774 \AA\ \OI\, line and \CaII\ NIR triplet soon after the peak spectra. The lack of high velocity Ca/O/Fe prevents the resulting features from blending strongly as seen in the He-poor/rich mixed He core models. The high energy mixed CO core models with 0.2/0.5/1.0 \msun\ of attached He shells contain too little Ca/O/Fe to produce high degrees of blending that would result in a broad lined Type Ib analogue to the Type Ic-BL. If we do completely mix the ejecta including the entire H/He shell, as seen in the He-poor/rich mixed He core models, the placement of high velocity O and Ca does result in blending at 8 foes and a gradual increase in strength of \HeI\ lines. However, the lack of Type Ib SNe with strongly blended \CaII\ features as seen in Section \ref{sec:obs_Ib_sne} does not favor this result.

Despite the high energy of the 5 and 8 foe He-poor/rich mixed CO core models in Figs \ref{fig:allhe_full} and \ref{fig:allhe_comp}, the \CaII\ NIR triplet and 7774 \AA\ \OI\ line remain unblended. In the He-poor/rich mixed He core models at 5 and 8 foes, these lines are blended for most epochs. This suggests that a high \ek\ alone is not enough to produce significant blending of spectral features. A physical mixing mechanism, as needed to mix Ca/O/Fe to high velocities in order to produce strong blending of their respective lines, could be by strong jets or very asymmetric explosions. Events showing evidence for this, such as SN1998bw \citep{Mazzali_2001} or SN1997ef \citep{Iwamoto_2000,Mazzali_2000}, are often correlated with high explosion energies.

\citet{10.1093/mnras/staa123} showed that mixed CO cores (called fully mixed in that work) was able to approximately replicate several epochs of several supernovae. For Type Ic events, the CO-rich core often requires some amount of mixing or deviation from standard explosion models when these events are modelled in order to reproduce spectra and light curves such as in SN1994I \citep{2006MNRAS.369.1939S}. It is unlikely to see a completely mixed He core with high masses of He. Despite this, the mixed He core models do replicate some observed behaviours, such as the \HeI\ lines growing in strength over time. Depending on \ek\ and \mhe, the models with a fully mixed CO core produce early optical \HeI\ lines before being blended into nearby features. 

The \HeI\ lines in the He--poor/rich models with a mixed CO core increase in width as \ek\ increases from 1 to 8 foes as seen in Figure \ref{fig:allhe_full}. The \HeI\ line in SN\,2009jf shown in Figures \ref{fig:early_epoch} to \ref{fig:later_epoch}, evolves from a fairly broad \HeI\ line early to a fairly narrow \HeI\ line later. As \ek\ increases, the velocities in which He is dominant increase as shown in Section \ref{subsec:abu_prof}, leading to broadened lines. For each \ek, the \HeI\ lines also narrow as the spectra evolve from pre to post-peak. 

\subsection{Type Ib-like Models}
\label{subsec:class_hepoorrich}

Figures \ref{fig:early_epoch} to \ref{fig:later_epoch} show that a combination of mixing and energy can reproduce the bulk spectral features of a sample of SN\,Ib over multiple epochs. The synthetic spectra and SN\,Ic comparisons in \citet{10.1093/mnras/staa123} showed that near--peak spectra for observed SNe show similarity to a wider range of the synthetic spectra at the full range of \ek. The observed SNe\,Ib in this work show less diversity, with SNe\,2004gq and 2007uy showing similarities to the higher \ek\ 3 and 5 foe models but none showing strong similarities to the 8 foe models. The broad-lined SN\,Ic SN\,2016coi is included for comparison and shows some similarity to the 5 and 8 foe models despite the 1.0 \msun\ of He in these models. SNe\,2007Y, 2007C, and 2009jf were reasonably well reproduced with the 1 foe models. The estimated \ek\ for SN\,2007uy given in \citet{2013MNRAS.434.2032R} is $\sim 15$\,foe, while using the estimate given by Figures \ref{fig:early_epoch} to \ref{fig:later_epoch} suggests a much lower \ek\ of 3-5 foes. SN\,2007Y is thought to have \ek\,$< 1$ foe and is very similar to the 1 foe models with both mixing approximations with a fairly consistent shift in line velocity. The overly mixed He cores blend the \CaII\ and \OI\ line that is not reflected in these observed events.

\subsection{SN\,IIb-like Models}
\label{subsec:class_hepoorrich}

For the SN\,IIb comparisons in Figures \ref{fig:IIb_early_epoch} to \ref{fig:IIb_later_epoch}, six SNe were considered and only SN\,2003bg did not fit best between the 1 foe models with both mixing approximations at the top of the three figures. The SN\,IIb observed and modelled parameter space is obviously narrower than both the SN\,Ib and Ic parameter spaces \citep{doi:10.1093/mnras/stw299}. The evolutionary paths, binary or single star evolution, capable of producing a progenitor that would result in a SN\,IIb  could be limited by mass, such that increasing or decreasing the progenitor mass would result in H-free SNe\,Ib/c or H-rich SNe\,II-P/L. This is partially backed up by the modelled \mej\ parameter space favoring lighter masses in a narrower range than for other SE--SNe.



\subsection{Helium Mass in Type Ib/c}
\label{subsec:class_hepoorrich}


For the He-poor/rich mixed CO core models, the early formation of \OI\,7774  places the photosphere in or very near the CO rich core. The composition above this region is comprised of $\sim$\,0.2--1.0 \msun\ of He, depending on the stripping. The \HeI\,6678 and 7065 lines are easily observable in the 1 and 3 foe models, but line broadening makes them difficult to see in the 5 and 8 foe models. Narrow-lined spectra in the lower energy models make separate identification of spectral features that share a common wavelength range, like \SiII\ and \Ha\ or \HeI\ and \NaI\,D, much easier. The addition of line blending of weak features can make early detections of \HeI\ or \Ha\ much more difficult. The mixed CO core model with 0.2 \msun\ of He (Fig. \ref{fig:allhe_full}(d)) shows this behaviour the strongest, as the 6678 \AA\ region is dominated by \SiII\ and this line changes only slightly throughout the evolution of the spectra. Without observing the NIR \HeI\ lines, this may be classified as a SN\,Ic despite the mass of He present in the model. If we consider \HeI\,20581 \AA\, in Figures \ref{fig:allhe_full} and \ref{fig:allhe_comp}, then identification of He in these models is easier. This line is always visible for the He-poor/rich mixed CO core models and sometimes visible for the He-poor/rich mixed He core models. 

\begin{figure}
    \centering
    \includegraphics[scale=.55]{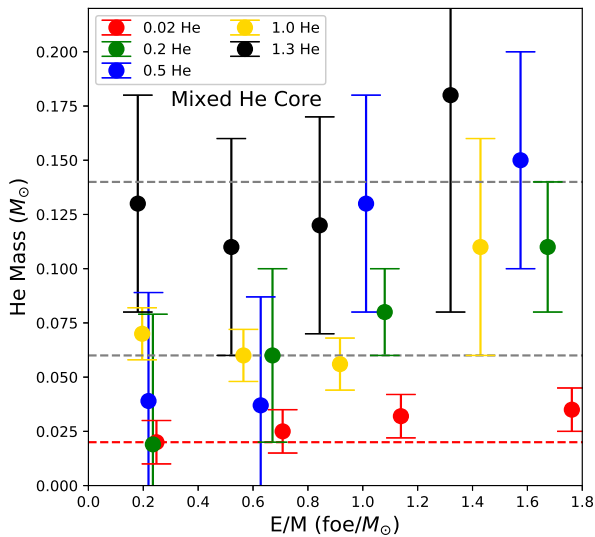}
    \caption{The He mass in the line forming region when the mixed He core models show all three optical and both NIR \HeI\ lines measured at the earliest epoch with error estimates based upon the masses in the nearby epochs. The dashed grey lines are the estimated ranges given in \citet{2012MNRAS.422...70H} and the red dashed line is if one only considers NIR features  \label{fig:min_he_mass}}
\end{figure}

Figure \ref{fig:min_he_mass} shows that the relatively little He mass can be hidden in the mixed He core models shown in this work.  The mixed CO core models always show the \HeI\ lines and have $\sim$0.1--0.5 \msun\ of He in the line forming region of the earliest epochs. There is a weak correlation that as the \eom\ ratio increases, up to $\sim$0.18 \msun\ He that can be hidden before the optical \HeI\ lines become visible. This ``hideable'' helium drops to 0.02 \msun\ of He or less (the red dashed line) if we only consider the NIR lines, as shown in the bare CO core models for example. Even if we show no optical \HeI\ in the earliest spectra, the lines still form in the later spectra as shown in Fig. \ref{fig:cocore_8foe}, meaning the He is not truly hidden.

The models in this work contain no sodium and therefore show no \NaI\,D line. For the bare CO cores models, the mixing approximation would place sodium throughout the ejecta, very likely leading to a strong feature at all epochs. The primary optical \HeI\ line in these models is weak or absent, but would be difficult or impossible to distinguish if \NaI\,D is present. This requires the use of other \HeI\ lines for strong identification as previously mentioned. Only considering optical \HeI\ lines, the two SNe in Figure \ref{fig:cocore_8foe} are similar until the final two epochs but deviate quickly in the NIR.

\subsection{Limitations}\label{subsec:limitations}

The plots shown in Sections \ref{sec:obs_Ib_sne} and \ref{sec:obs_IIb_sne} all use scaled flux to compare observed events to our synthetic spectra. Scaling the spectra hides the absolute flux of the SNe, which is determined by the luminosity of the event. This luminosity combined with a photospheric velocity sets the radiative temperature above the photosphere. Combining the radiative temperature with the abundances sets the excitation and ionization state of the gas. All of the above factors determine electron and line opacity, resulting in a final spectrum. A change in absolute flux changes the luminosity and can drastically alter the observed spectra \citep{Ashall2020}. Giving a proper estimate for an observed SN requires fitting both the observed light curve and the spectra using abundance tomography \citep{2005MNRAS.360.1231S}. As such, using the scaled flux is a reasonable approximation to determine possible line identifications and ejecta energy, but not absolute line strength or several other observational quantities.

For the Type Ic SNe in \citet{10.1093/mnras/staa123}, completely mixed is a reasonable approximation due to the relative size and compactness of CO cores. This may be less likely in He--rich SNe. The behaviour of the \CaII\ triplet and \OI\ line in the mixed He core models rarely matched the observed behaviours of these features. The mixed He core approximation places all elements across the entire ejecta, including in to the He--rich atmosphere. The difference in looking at a mixed He or mixed CO core is better framed considering regions in which the ejecta shows mixing. SN\,2007C in Section \ref{subsec:2007c} is a good example in which the best matching synthetic spectra require different mixing approximations as the epoch increases. As such, a mixed He core model is a good estimate of some spectral behaviours and identifications, but not line width or strength.

The above \HeI\ identifications all had a \HeI\ dominated 5875 \AA\ region which allowed us to track its development. If we include sodium, which is a valid expectation in CO core models, a clean 5875 \AA\ feature is likely not to occur. For weak optical \HeI\ lines in Fig. \ref{fig:cocore_complete}, a strong or intermediate \NaI\,D line would likely be dominant. Using line velocities, such as in \citet{2018A&A...618A..37F}, may shed light on how these features interact but solely relying on this single line can result in incorrect assumptions. This can be seen in Fig. \ref{fig:cocore_8foe}, in which one could argue that the 8 foe mixed He core model with 1.0 \msun\ of He is hiding 0.2--1.0 \msun\ of He based upon optical lines only. Using the NIR spectra, this is clearly not the case, showing easily observable \HeI\ features, and the final epochs show a much stronger optical \HeI\ lines that were not present in the earliest epochs. SN2016coi, in Figures \ref{fig:early_epoch} to \ref{fig:later_epoch}, shows evidence of similarity mixed He core model with 1.0 \msun\ of He but this similarity is lost in the later epochs as the He line continues to grow in strength.
\section{Conclusion} 
\label{sec:conc}

In this work we exploded a 22\,\msun\ progenitor that was artificially stripped of some or all of its H/He prior to core collapse at an \ek\ of 1, 3, 5, and 8 foes using the 1, but is likely not observable given the extreme mixing-D explosion code \textsc{Prometheus-HOTB}. For each stripped model at each \ek\ we applied three mixing approximations that represent a mixed He core, mixed CO core with an unmixed H/He shell, and ejecta without mixing. We generated $\sim$7 spectra for each combination of stripping, \ek, and mixing, covering $\sim$4 weeks of spectral evolution. We compared the synthetic spectra to a set of observed SNe\,Ib and IIb.

We showed that the 8 foe mixed He core model with 1.0\,\msun\ of He has no optical \HeI\ lines in the early spectra, but does show stronger NIR \HeI\ lines. This shows that for finding trace amounts He in SNe, one requires the use of both optical and NIR spectra. Relying on the primary optical \HeI\ line is complicated by the presence of the nearby strong \NaI\,D line which we did not have in the models in this work. In addition, if He is in the model, \HeI\ lines will almost always be observable in the NIR, but in the optical region, identification is complicated by mixing, \ek, and abundances.

We found that for the models in this work, one cannot hide a significant mass of \HeI\ without line formation at some epoch or at some wavelength. The low mass of He in the bare CO core models at 8 foes still produced NIR \HeI\ lines, but not strong optical \HeI\ lines. As the mass of He increases in the He--poor/rich models, the visibility of the \HeI\ lines became easier, but can be complicated at higher explosion energies For He identification in SE--SNE, one needs to consider both optical and NIR spectra as even low masses of He can produce saturated NIR lines.

We found that the He--poor/rich models with low to moderate \ek\ can replicate multiple spectral epochs of a variety of observed SNe\,Ib. This works particularly well when observed and modelled \ek\, are similar, suggesting that this can be a coarse method to estimate the \ek. Few of our H--poor/He--rich models showed strong similarity to that of observed Type IIb SNe. The observed \ek\ parameter space for SNe\,Ib and IIb is fairly narrow and is populated with low energy events, such that the high energy He--poor/rich and H--poor/He--rich models with significant mixing matched few observed SNe. 

Future work will complete this study by including a higher and lower mass progenitor model stripped and exploded at the same \ek\, values, as well as lower than 1 foe for low mass events and higher than 8 foe for higher mass events. Thus the \eom\ ratios cover a wide range from 0.25 to 2 foe/\msun\ or higher. The resulting grid of models and spectra will cover a wide \mej\ range of radioactively powered SE--SNe as well as a wide range of spectral epochs which will constrain the parameter space for SE--SNe.

\section*{ACKNOWLEDGMENTS}
At Garching, this project was supported by the European Research Council through grant ERC-AdG No.341157-COCO2CASA, and by the Deutsche Forschungsgemeinschaft through Sonderforschungbereich SFB 1258 `Neutrinos and Dark Matter in Astro- and Particle Physics' (NDM) and the Excellence Cluster Universe 'ORIGINS: From the Origin of the Universe to the First Building Blocks of Life' (EXC 2094 --- 390783311; \url{https://www.origins-cluster.de})

\bibliographystyle{mnras}
\bibliography{ref_list_typeib}

\begin{thebibliography}{}
\makeatletter
\relax
\def\mn@urlcharsother{\let\do\@makeother \do\$\do\&\do\#\do\^\do\_\do\%\do\~}
\def\mn@doi{\begingroup\mn@urlcharsother \@ifnextchar [ {\mn@doi@}
  {\mn@doi@[]}}
\def\mn@doi@[#1]#2{\def\@tempa{#1}\ifx\@tempa\@empty \href
  {http://dx.doi.org/#2} {doi:#2}\else \href {http://dx.doi.org/#2} {#1}\fi
  \endgroup}
\def\mn@eprint#1#2{\mn@eprint@#1:#2::\@nil}
\def\mn@eprint@arXiv#1{\href {http://arxiv.org/abs/#1} {{\tt arXiv:#1}}}
\def\mn@eprint@dblp#1{\href {http://dblp.uni-trier.de/rec/bibtex/#1.xml}
  {dblp:#1}}
\def\mn@eprint@#1:#2:#3:#4\@nil{\def\@tempa {#1}\def\@tempb {#2}\def\@tempc
  {#3}\ifx \@tempc \@empty \let \@tempc \@tempb \let \@tempb \@tempa \fi \ifx
  \@tempb \@empty \def\@tempb {arXiv}\fi \@ifundefined
  {mn@eprint@\@tempb}{\@tempb:\@tempc}{\expandafter \expandafter \csname
  mn@eprint@\@tempb\endcsname \expandafter{\@tempc}}}

\bibitem[\protect\citeauthoryear{{Arnett}}{{Arnett}}{1982}]{1982ApJ...253..785A}
{Arnett} W.~D.,  1982, \mn@doi [\apj] {10.1086/159681}, \href
  {http://adsabs.harvard.edu/abs/1982ApJ...253..785A} {253, 785}

\bibitem[\protect\citeauthoryear{Ashall \& Mazzali}{Ashall \&
  Mazzali}{2020}]{Ashall2020}
Ashall C.,  Mazzali P.~A.,  2020, \mn@doi [\mnras] {10.1093/mnras/staa212},
  492, 5956

\bibitem[\protect\citeauthoryear{Ashall et~al.,}{Ashall
  et~al.}{2019}]{10.1093/mnras/stz1588}
Ashall C.,  et~al., 2019, \mn@doi [\mnras] {10.1093/mnras/stz1588}

\bibitem[\protect\citeauthoryear{{Axelrod}}{{Axelrod}}{1980}]{1980PhDT.........1A}
{Axelrod} T.~S.,  1980, PhD thesis, California Univ., Santa Cruz.

\bibitem[\protect\citeauthoryear{{Benetti} et~al.,}{{Benetti}
  et~al.}{2011}]{2011MNRAS.411.2726B}
{Benetti} S.,  et~al., 2011, \mn@doi [\mnras]
  {10.1111/j.1365-2966.2010.17873.x}, \href
  {https://ui.adsabs.harvard.edu/abs/2011MNRAS.411.2726B} {411, 2726}

\bibitem[\protect\citeauthoryear{{Bersten} et~al.,}{{Bersten}
  et~al.}{2012}]{2012ApJ...757...31B}
{Bersten} M.~C.,  et~al., 2012, \mn@doi [\apj] {10.1088/0004-637X/757/1/31},
  \href {https://ui.adsabs.harvard.edu/abs/2012ApJ...757...31B} {757, 31}

\bibitem[\protect\citeauthoryear{Branch et~al.,}{Branch
  et~al.}{2002}]{Branch_2002}
Branch D.,  et~al., 2002, \mn@doi [\apj] {10.1086/338127}, 566, 1005

\bibitem[\protect\citeauthoryear{{Burrows}, {Hayes}  \& {Fryxell}}{{Burrows}
  et~al.}{1995}]{1995ApJ...450..830B}
{Burrows} A.,  {Hayes} J.,   {Fryxell} B.~A.,  1995, \mn@doi [\apj]
  {10.1086/176188}, \href {http://adsabs.harvard.edu/abs/1995ApJ...450..830B}
  {450, 830}

\bibitem[\protect\citeauthoryear{Cappellaro, Mazzali, Benetti, Danziger,
  Turatto, della Valle  \& Patat}{Cappellaro
  et~al.}{1997}]{1997A&A...328..203C}
Cappellaro E.,  Mazzali P.~A.,  Benetti S.,  Danziger I.~J.,  Turatto M.,
  della Valle M.,   Patat F.,  1997, \mn@doi [\aap] {1997A
  \href {http://ukads.nottingham.ac.uk/abs/1997A%26A...328..203C} {328, 203}

\bibitem[\protect\citeauthoryear{{Chornock} et~al.,}{{Chornock}
  et~al.}{2011}]{2011ApJ...739...41C}
{Chornock} R.,  et~al., 2011, \mn@doi [\apj] {10.1088/0004-637X/739/1/41},
  \href {http://adsabs.harvard.edu/abs/2011ApJ...739...41C} {739, 41}

\bibitem[\protect\citeauthoryear{{Drout} et~al.,}{{Drout}
  et~al.}{2011}]{2011ApJ...741...97D}
{Drout} M.~R.,  et~al., 2011, \mn@doi [\apj] {10.1088/0004-637X/741/2/97},
  \href {https://ui.adsabs.harvard.edu/abs/2011ApJ...741...97D} {741, 97}

\bibitem[\protect\citeauthoryear{{Drout} et~al.,}{{Drout}
  et~al.}{2016}]{Drout2016}
{Drout} M.~R.,  et~al., 2016, \mn@doi [\apj] {10.3847/0004-637X/821/1/57},
  \href {http://adsabs.harvard.edu/abs/2016ApJ...821...57D} {821, 57}

\bibitem[\protect\citeauthoryear{{Eastman}, {Woosley}, {Weaver}  \&
  {Pinto}}{{Eastman} et~al.}{1994}]{1994ApJ...430..300E}
{Eastman} R.~G.,  {Woosley} S.~E.,  {Weaver} T.~A.,   {Pinto} P.~A.,  1994,
  \mn@doi [\apj] {10.1086/174404}, \href
  {http://adsabs.harvard.edu/abs/1994ApJ...430..300E} {430, 300}

\bibitem[\protect\citeauthoryear{{Elmhamdi}, {Danziger}, {Branch},
  {Leibundgut}, {Baron}  \& {Kirshner}}{{Elmhamdi} et~al.}{2006}]{Elmhamdi2006}
{Elmhamdi} A.,  {Danziger} I.~J.,  {Branch} D.,  {Leibundgut} B.,  {Baron} E.,
   {Kirshner} R.~P.,  2006, \mn@doi [\aap] {10.1051/0004-6361:20054366}, \href
  {http://adsabs.harvard.edu/abs/2006A%26A...450..305E} {450, 305}

\bibitem[\protect\citeauthoryear{{Ergon, M.} et~al.,}{{Ergon, M.}
  et~al.}{2015}]{ergon2015}
{Ergon, M.} et~al., 2015, \mn@doi [A\&A] {10.1051/0004-6361/201424592}, 580,
  A142

\bibitem[\protect\citeauthoryear{{Ertl}, {Janka}, {Woosley}, {Sukhbold}  \&
  {Ugliano}}{{Ertl} et~al.}{2016}]{ertl2016}
{Ertl} T.,  {Janka} H.~T.,  {Woosley} S.~E.,  {Sukhbold} T.,   {Ugliano} M.,
  2016, \mn@doi [\apj] {10.3847/0004-637X/818/2/124}, \href
  {https://ui.adsabs.harvard.edu/abs/2016ApJ...818..124E} {818, 124}

\bibitem[\protect\citeauthoryear{{Ertl}, {Woosley}, {Sukhbold}  \&
  {Janka}}{{Ertl} et~al.}{2020}]{ertl2020}
{Ertl} T.,  {Woosley} S.~E.,  {Sukhbold} T.,   {Janka} H.~T.,  2020, \mn@doi
  [\apj] {10.3847/1538-4357/ab6458}, \href
  {https://ui.adsabs.harvard.edu/abs/2020ApJ...890...51E} {890, 51}

\bibitem[\protect\citeauthoryear{{Filippenko}}{{Filippenko}}{1997}]{1997ARA&A..35..309F}
{Filippenko} A.~V.,  1997, \mn@doi [\araa] {10.1146/annurev.astro.35.1.309},
  \href {http://ukads.nottingham.ac.uk/abs/1997ARA%26A..35..309F} {35, 309}

\bibitem[\protect\citeauthoryear{{Filippenko} et~al.,}{{Filippenko}
  et~al.}{1995}]{1995ApJ...450L..11F}
{Filippenko} A.~V.,  et~al., 1995, \mn@doi [\apjl] {10.1086/309659}, \href
  {http://adsabs.harvard.edu/abs/1995ApJ...450L..11F} {450, L11}

\bibitem[\protect\citeauthoryear{Folatelli et~al.,}{Folatelli
  et~al.}{2014}]{Folatelli_2014}
Folatelli G.,  et~al., 2014, \mn@doi [\apj] {10.1088/0004-637x/792/1/7}, 792, 7

\bibitem[\protect\citeauthoryear{{Fremling} et~al.,}{{Fremling}
  et~al.}{2018}]{2018A&A...618A..37F}
{Fremling} C.,  et~al., 2018, \mn@doi [\aap] {10.1051/0004-6361/201731701},
  \href {https://ui.adsabs.harvard.edu/abs/2018A&A...618A..37F} {618, A37}

\bibitem[\protect\citeauthoryear{{Graham}}{{Graham}}{1988}]{1988ApJ...335L..53G}
{Graham} J.~R.,  1988, \mn@doi [\apjl] {10.1086/185338}, \href
  {https://ui.adsabs.harvard.edu/abs/1988ApJ...335L..53G} {335, L53}

\bibitem[\protect\citeauthoryear{{Hachinger}, {Mazzali}, {Taubenberger},
  {Hillebrandt}, {Nomoto}  \& {Sauer}}{{Hachinger}
  et~al.}{2012}]{2012MNRAS.422...70H}
{Hachinger} S.,  {Mazzali} P.~A.,  {Taubenberger} S.,  {Hillebrandt} W.,
  {Nomoto} K.,   {Sauer} D.~N.,  2012, \mn@doi [\mnras]
  {10.1111/j.1365-2966.2012.20464.x}, \href
  {http://adsabs.harvard.edu/abs/2012MNRAS.422...70H} {422, 70}

\bibitem[\protect\citeauthoryear{{Hamuy} et~al.,}{{Hamuy}
  et~al.}{2009}]{2009ApJ...703.1612H}
{Hamuy} M.,  et~al., 2009, \mn@doi [\apj] {10.1088/0004-637X/703/2/1612}, \href
  {https://ui.adsabs.harvard.edu/abs/2009ApJ...703.1612H} {703, 1612}

\bibitem[\protect\citeauthoryear{{Harkness} et~al.,}{{Harkness}
  et~al.}{1987}]{1987ApJ...317..355H}
{Harkness} R.~P.,  et~al., 1987, \mn@doi [\apj] {10.1086/165283}, \href
  {https://ui.adsabs.harvard.edu/abs/1987ApJ...317..355H} {317, 355}

\bibitem[\protect\citeauthoryear{{Iwamoto} et~al.,}{{Iwamoto}
  et~al.}{1998}]{1998Natur.395..672I}
{Iwamoto} K.,  et~al., 1998, \mn@doi [\nat] {10.1038/27155}, \href
  {http://adsabs.harvard.edu/abs/1998Natur.395..672I} {395, 672}

\bibitem[\protect\citeauthoryear{Iwamoto et~al.,}{Iwamoto
  et~al.}{2000}]{Iwamoto_2000}
Iwamoto K.,  et~al., 2000, \mn@doi [\apj] {10.1086/308761}, 534, 660

\bibitem[\protect\citeauthoryear{{Janka} \& {M{\"u}ller}}{{Janka} \&
  {M{\"u}ller}}{1996}]{1996A&A...306..167J}
{Janka} H.-T.,  {M{\"u}ller} E.,  1996, \aap, \href
  {http://adsabs.harvard.edu/abs/1996A%26A...306..167J} {306, 167}

\bibitem[\protect\citeauthoryear{{Lucy}}{{Lucy}}{1991}]{1991ApJ...383..308L}
{Lucy} L.~B.,  1991, \mn@doi [\apj] {10.1086/170787}, \href
  {https://ui.adsabs.harvard.edu/abs/1991ApJ...383..308L} {383, 308}

\bibitem[\protect\citeauthoryear{{Lucy}}{{Lucy}}{1999}]{1999A&A...345..211L}
{Lucy} L.~B.,  1999, \aap, \href
  {http://adsabs.harvard.edu/abs/1999A%26A...345..211L} {345, 211}

\bibitem[\protect\citeauthoryear{{Mazzali}}{{Mazzali}}{2000}]{2000A&A...363..705M}
{Mazzali} P.~A.,  2000, \aap, \href
  {http://adsabs.harvard.edu/abs/2000A%26A...363..705M} {363, 705}

\bibitem[\protect\citeauthoryear{{Mazzali} \& {Lucy}}{{Mazzali} \&
  {Lucy}}{1993}]{1993A&A...279..447M}
{Mazzali} P.~A.,  {Lucy} L.~B.,  1993, \aap, \href
  {https://ui.adsabs.harvard.edu/abs/1993A&A...279..447M} {279, 447}

\bibitem[\protect\citeauthoryear{Mazzali, Iwamoto  \& Nomoto}{Mazzali
  et~al.}{2000}]{Mazzali_2000}
Mazzali P.~A.,  Iwamoto K.,   Nomoto K.,  2000, \mn@doi [\apj]
  {10.1086/317808}, 545, 407

\bibitem[\protect\citeauthoryear{{Mazzali}, {Nomoto}, {Cappellaro}, {Nakamura},
  {Umeda}  \& {Iwamoto}}{{Mazzali} et~al.}{2001a}]{2001ApJ...547..988M}
{Mazzali} P.~A.,  {Nomoto} K.,  {Cappellaro} E.,  {Nakamura} T.,  {Umeda} H.,
  {Iwamoto} K.,  2001a, \mn@doi [\apj] {10.1086/318428}, \href
  {https://ui.adsabs.harvard.edu/abs/2001ApJ...547..988M} {547, 988}

\bibitem[\protect\citeauthoryear{Mazzali, Nomoto, Patat  \& Maeda}{Mazzali
  et~al.}{2001b}]{Mazzali_2001}
Mazzali P.~A.,  Nomoto K.,  Patat F.,   Maeda K.,  2001b, \mn@doi [\apj]
  {10.1086/322420}, 559, 1047

\bibitem[\protect\citeauthoryear{{Mazzali} et~al.,}{{Mazzali}
  et~al.}{2002}]{2002ApJ...572L..61M}
{Mazzali} P.~A.,  et~al., 2002, \mn@doi [\apjl] {10.1086/341504}, \href
  {http://adsabs.harvard.edu/abs/2002ApJ...572L..61M} {572, L61}

\bibitem[\protect\citeauthoryear{{Mazzali}, {Deng}, {Maeda}, {Nomoto},
  {Filippenko}  \& {Matheson}}{{Mazzali} et~al.}{2004}]{2004ApJ...614..858M}
{Mazzali} P.~A.,  {Deng} J.,  {Maeda} K.,  {Nomoto} K.,  {Filippenko} A.~V.,
  {Matheson} T.,  2004, \mn@doi [\apj] {10.1086/423888}, \href
  {https://ui.adsabs.harvard.edu/abs/2004ApJ...614..858M} {614, 858}

\bibitem[\protect\citeauthoryear{Mazzali et~al.,}{Mazzali
  et~al.}{2005}]{Mazzali1284}
Mazzali P.~A.,  et~al., 2005, \mn@doi [\sci] {10.1126/science.1111384}, 308,
  1284

\bibitem[\protect\citeauthoryear{{Mazzali}, {Deng}, {Hamuy}  \&
  {Nomoto}}{{Mazzali} et~al.}{2009}]{2009ApJ...703.1624M}
{Mazzali} P.~A.,  {Deng} J.,  {Hamuy} M.,   {Nomoto} K.,  2009, \mn@doi [\apj]
  {10.1088/0004-637X/703/2/1624}, \href
  {https://ui.adsabs.harvard.edu/abs/2009ApJ...703.1624M} {703, 1624}

\bibitem[\protect\citeauthoryear{{Modjaz} et~al.,}{{Modjaz}
  et~al.}{2014}]{2014AJ....147...99M}
{Modjaz} M.,  et~al., 2014, \mn@doi [\aj] {10.1088/0004-6256/147/5/99}, \href
  {http://adsabs.harvard.edu/abs/2014AJ....147...99M} {147, 99}

\bibitem[\protect\citeauthoryear{Morales-Garoffolo et~al.,}{Morales-Garoffolo
  et~al.}{2015}]{10.1093/mnras/stv1972}
Morales-Garoffolo A.,  et~al., 2015, \mn@doi [\mnras] {10.1093/mnras/stv1972},
  454, 95

\bibitem[\protect\citeauthoryear{{Nomoto}, {Suzuki}, {Shigeyama}, {Kumagai},
  {Yamaoka}  \& {Saio}}{{Nomoto} et~al.}{1993}]{1993Natur.364..507N}
{Nomoto} K.,  {Suzuki} T.,  {Shigeyama} T.,  {Kumagai} S.,  {Yamaoka} H.,
  {Saio} H.,  1993, \mn@doi [\nat] {10.1038/364507a0}, \href
  {https://ui.adsabs.harvard.edu/abs/1993Natur.364..507N} {364, 507}

\bibitem[\protect\citeauthoryear{Nomoto, Yamaoka, Pols, van~den Heuvel,
  Iwamoto, Kumagai  \& Shigeyama}{Nomoto et~al.}{1994}]{Nomoto1994}
Nomoto K.,  Yamaoka H.,  Pols O.~R.,  van~den Heuvel E. P.~J.,  Iwamoto K.,
  Kumagai S.,   Shigeyama T.,  1994, \mn@doi [\nat] {10.1038/371227a0}, 371,
  227

\bibitem[\protect\citeauthoryear{Nomoto, Iwamoto  \& Suzuki}{Nomoto
  et~al.}{1995}]{NOMOTO1995173}
Nomoto K.,  Iwamoto K.,   Suzuki T.,  1995, \mn@doi [Phys. Rep.]
  {https://doi.org/10.1016/0370-1573(94)00107-E}, 256, 173

\bibitem[\protect\citeauthoryear{Prentice \& Mazzali}{Prentice \&
  Mazzali}{2017}]{10.1093/mnras/stx980}
Prentice S.~J.,  Mazzali P.~A.,  2017, \mn@doi [\mnras] {10.1093/mnras/stx980},
  469, 2672

\bibitem[\protect\citeauthoryear{Prentice et~al.,}{Prentice
  et~al.}{2016}]{doi:10.1093/mnras/stw299}
Prentice S.~J.,  et~al., 2016, \mn@doi [\mnras] {10.1093/mnras/stw299}, 458,
  2973

\bibitem[\protect\citeauthoryear{{Prentice} et~al.,}{{Prentice}
  et~al.}{2018a}]{2018MNRAS.478.4162P}
{Prentice} S.~J.,  et~al., 2018a, \mn@doi [\mnras] {10.1093/mnras/sty1223},
  \href {http://adsabs.harvard.edu/abs/2018MNRAS.478.4162P} {478, 4162}

\bibitem[\protect\citeauthoryear{Prentice et~al.,}{Prentice
  et~al.}{2018b}]{10.1093/mnras/sty3399}
Prentice S.~J.,  et~al., 2018b, \mn@doi [\mnras] {10.1093/mnras/sty3399}, 485,
  1559

\bibitem[\protect\citeauthoryear{{Roy} et~al.,}{{Roy}
  et~al.}{2013}]{2013MNRAS.434.2032R}
{Roy} R.,  et~al., 2013, \mn@doi [\mnras] {10.1093/mnras/stt1148}, \href
  {https://ui.adsabs.harvard.edu/abs/2013MNRAS.434.2032R} {434, 2032}

\bibitem[\protect\citeauthoryear{{Sahu}, {Gurugubelli}, {Anupama}  \&
  {Nomoto}}{{Sahu} et~al.}{2011}]{2011MNRAS.413.2583S}
{Sahu} D.~K.,  {Gurugubelli} U.~K.,  {Anupama} G.~C.,   {Nomoto} K.,  2011,
  \mn@doi [\mnras] {10.1111/j.1365-2966.2011.18326.x}, \href
  {https://ui.adsabs.harvard.edu/abs/2011MNRAS.413.2583S} {413, 2583}

\bibitem[\protect\citeauthoryear{{Sauer}, {Mazzali}, {Deng}, {Valenti},
  {Nomoto}  \& {Filippenko}}{{Sauer} et~al.}{2006}]{2006MNRAS.369.1939S}
{Sauer} D.~N.,  {Mazzali} P.~A.,  {Deng} J.,  {Valenti} S.,  {Nomoto} K.,
  {Filippenko} A.~V.,  2006, \mn@doi [\mnras]
  {10.1111/j.1365-2966.2006.10438.x}, \href
  {http://adsabs.harvard.edu/abs/2006MNRAS.369.1939S} {369, 1939}

\bibitem[\protect\citeauthoryear{{Shigeyama}, {Nomoto}, {Tsujimoto}  \&
  {Hashimoto}}{{Shigeyama} et~al.}{1990}]{1990ApJ...361L..23S}
{Shigeyama} T.,  {Nomoto} K.,  {Tsujimoto} T.,   {Hashimoto} M.-A.,  1990,
  \mn@doi [\apjl] {10.1086/185818}, \href
  {https://ui.adsabs.harvard.edu/abs/1990ApJ...361L..23S} {361, L23}

\bibitem[\protect\citeauthoryear{Shivvers et~al.,}{Shivvers
  et~al.}{2018}]{10.1093/mnras/sty2719}
Shivvers I.,  et~al., 2018, \mn@doi [\mnras] {10.1093/mnras/sty2719}, 482, 1545

\bibitem[\protect\citeauthoryear{{Smith}, {Gehrz}, {Hinz}, {Hoffmann}, {Hora},
  {Mamajek}  \& {Meyer}}{{Smith} et~al.}{2003}]{Smith2003}
{Smith} N.,  {Gehrz} R.~D.,  {Hinz} P.~M.,  {Hoffmann} W.~F.,  {Hora} J.~L.,
  {Mamajek} E.~E.,   {Meyer} M.~R.,  2003, \mn@doi [\aj] {10.1086/346278},
  \href {http://adsabs.harvard.edu/abs/2003AJ....125.1458S} {125, 1458}

\bibitem[\protect\citeauthoryear{{Stehle}, {Mazzali}, {Benetti}  \&
  {Hillebrandt}}{{Stehle} et~al.}{2005}]{2005MNRAS.360.1231S}
{Stehle} M.,  {Mazzali} P.~A.,  {Benetti} S.,   {Hillebrandt} W.,  2005,
  \mn@doi [\mnras] {10.1111/j.1365-2966.2005.09116.x}, \href
  {http://ukads.nottingham.ac.uk/abs/2005MNRAS.360.1231S} {360, 1231}

\bibitem[\protect\citeauthoryear{{Stevance} et~al.,}{{Stevance}
  et~al.}{2017}]{2017MNRAS.469.1897S}
{Stevance} H.~F.,  et~al., 2017, \mn@doi [\mnras] {10.1093/mnras/stx970}, \href
  {http://adsabs.harvard.edu/abs/2017MNRAS.469.1897S} {469, 1897}

\bibitem[\protect\citeauthoryear{{Stritzinger} et~al.,}{{Stritzinger}
  et~al.}{2009}]{2009ApJ...696..713S}
{Stritzinger} M.,  et~al., 2009, \mn@doi [\apj] {10.1088/0004-637X/696/1/713},
  \href {https://ui.adsabs.harvard.edu/abs/2009ApJ...696..713S} {696, 713}

\bibitem[\protect\citeauthoryear{{Sukhbold}, {Ertl}, {Woosley}, {Brown}  \&
  {Janka}}{{Sukhbold} et~al.}{2016}]{2016ApJ...821...38S}
{Sukhbold} T.,  {Ertl} T.,  {Woosley} S.~E.,  {Brown} J.~M.,   {Janka} H.~T.,
  2016, \mn@doi [\apj] {10.3847/0004-637X/821/1/38}, \href
  {https://ui.adsabs.harvard.edu/abs/2016ApJ...821...38S} {821, 38}

\bibitem[\protect\citeauthoryear{{Swartz}, {Clocchiatti}, {Benjamin}, {Lester}
  \& {Wheeler}}{{Swartz} et~al.}{1993}]{1993Natur.365..232S}
{Swartz} D.~A.,  {Clocchiatti} A.,  {Benjamin} R.,  {Lester} D.~F.,   {Wheeler}
  J.~C.,  1993, \mn@doi [\nat] {10.1038/365232a0}, \href
  {https://ui.adsabs.harvard.edu/abs/1993Natur.365..232S} {365, 232}

\bibitem[\protect\citeauthoryear{Tanaka et~al.,}{Tanaka
  et~al.}{2009}]{Tanaka_2009}
Tanaka M.,  et~al., 2009, \mn@doi [\apj] {10.1088/0004-637x/700/2/1680}, 700,
  1680

\bibitem[\protect\citeauthoryear{{Taubenberger} et~al.,}{{Taubenberger}
  et~al.}{2011}]{2011MNRAS.413.2140T}
{Taubenberger} S.,  et~al., 2011, \mn@doi [\mnras]
  {10.1111/j.1365-2966.2011.18287.x}, \href
  {https://ui.adsabs.harvard.edu/abs/2011MNRAS.413.2140T} {413, 2140}

\bibitem[\protect\citeauthoryear{Teffs, Ertl, Mazzali, Hachinger  \&
  Janka}{Teffs et~al.}{2020}]{10.1093/mnras/staa123}
Teffs J.,  Ertl T.,  Mazzali P.,  Hachinger S.,   Janka T.,  2020, \mn@doi
  [\mnras] {10.1093/mnras/staa123}, 492, 4369

\bibitem[\protect\citeauthoryear{{Ugliano}, {Janka}, {Marek}  \&
  {Arcones}}{{Ugliano} et~al.}{2012}]{2012ApJ...757...69U}
{Ugliano} M.,  {Janka} H.-T.,  {Marek} A.,   {Arcones} A.,  2012, \mn@doi
  [\apj] {10.1088/0004-637X/757/1/69}, \href
  {https://ui.adsabs.harvard.edu/abs/2012ApJ...757...69U} {757, 69}

\bibitem[\protect\citeauthoryear{{Valenti} et~al.,}{{Valenti}
  et~al.}{2011}]{2011MNRAS.416.3138V}
{Valenti} S.,  et~al., 2011, \mn@doi [\mnras]
  {10.1111/j.1365-2966.2011.19262.x}, \href
  {https://ui.adsabs.harvard.edu/abs/2011MNRAS.416.3138V} {416, 3138}

\bibitem[\protect\citeauthoryear{{Wang} \& {Wheeler}}{{Wang} \&
  {Wheeler}}{2008}]{2008ARA&A..46..433W}
{Wang} L.,  {Wheeler} J.~C.,  2008, \mn@doi [\araa]
  {10.1146/annurev.astro.46.060407.145139}, \href
  {https://ui.adsabs.harvard.edu/\#abs/2008ARA&A..46..433W} {46, 433}

\bibitem[\protect\citeauthoryear{{Woosley} \& {Weaver}}{{Woosley} \&
  {Weaver}}{1995}]{1995ApJS..101..181W}
{Woosley} S.~E.,  {Weaver} T.~A.,  1995, \mn@doi [\apjs] {10.1086/192237},
  \href {https://ui.adsabs.harvard.edu/abs/1995ApJS..101..181W} {101, 181}

\bibitem[\protect\citeauthoryear{Woosley, Heger  \& Weaver}{Woosley
  et~al.}{2002}]{RevModPhys.74.1015}
Woosley S.~E.,  Heger A.,   Weaver T.~A.,  2002, \mn@doi [Rev. Mod. Phys.]
  {10.1103/RevModPhys.74.1015}, 74, 1015

\bibitem[\protect\citeauthoryear{{Yaron} \& {Gal-Yam}}{{Yaron} \&
  {Gal-Yam}}{2012}]{2012PASP..124..668Y}
{Yaron} O.,  {Gal-Yam} A.,  2012, \mn@doi [\pasp] {10.1086/666656}, \href
  {http://adsabs.harvard.edu/abs/2012PASP..124..668Y} {124, 668}

\bibitem[\protect\citeauthoryear{Yoon, Woosley  \& Langer}{Yoon
  et~al.}{2010}]{Yoon_2010}
Yoon S.-C.,  Woosley S.~E.,   Langer N.,  2010, \mn@doi [\apj]
  {10.1088/0004-637x/725/1/940}, 725, 940

\bibitem[\protect\citeauthoryear{{Young}, {Baron}  \& {Branch}}{{Young}
  et~al.}{1995}]{1995ApJ...449L..51Y}
{Young} T.~R.,  {Baron} E.,   {Branch} D.,  1995, \mn@doi [\apjl]
  {10.1086/309618}, \href
  {https://ui.adsabs.harvard.edu/abs/1995ApJ...449L..51Y} {449, L51}

\makeatother
\end{thebibliography}



\appendix




\bsp	
\label{lastpage}
\end{document}